\documentclass[preprint,authoryear,5p,twocolumn,12pt]{elsarticle}

\usepackage{amsmath}
\usepackage{amssymb}
\usepackage{url}
\usepackage{multirow}
\usepackage{xcolor}
\newcommand{\degree}{\ensuremath{^\circ}}
\newcommand{\head}[1]{\textnormal{\textbf{#1}}}
\newcommand\T{\rule{0pt}{3ex}}       % Top strut
\newcommand\B{\rule[-1.5ex]{0pt}{0pt}} % Bottom strut
\journal{Icarus}

\begin{document}

\begin{frontmatter}



\title{The contamination of the surface of Vesta by impacts and the delivery of the dark material}


\author[iaps]{D. Turrini\corref{cor1}}
%%\ead{diego.turrini@iaps.inaf.it}
\author[bear]{J.-P. Combe}
\author[bear]{T. B. McCord}
\author[maxplank]{N. Oklay}
\author[maxplank]{J.-B Vincent}
\author[psi]{T.\ H. Prettyman}
\author[tennessee]{H.\ Y. McSween}
\author[vatican]{G. J. Consolmagno SJ}
\author[iaps]{M.\ C. De Sanctis}
\author[psi]{L. Le Corre}
\author[iaps]{A. Longobardo}
\author[iaps]{E. Palomba}
\author[ucla]{C.\ T. Russell}
\address[iaps]{Istituto di Astrofisica e Planetologia Spaziali INAF-IAPS, Via del Fosso del Cavaliere 100, 00133, Rome, Italy.}
\cortext[cor1]{Email address: diego.turrini@iaps.inaf.it}
\address[bear]{Bear Fight Institute, 22 Fiddler's Road, Box 667, Winthrop, Washington 98862, USA.}
\address[maxplank]{Max Planck Institute for Solar System Research
(MPS), Katlenburg-Lindau, Germany.}
\address[psi]{Planetary Science Institute, Tucson, Arizona 85719, USA.}
\address[tennessee]{University of Tennessee, Knoxville, Tennessee 37996, USA.}
\address[vatican]{Specola Vaticana, V-00120, Vatican City State.}
\address[ucla]{University of California, Los Angeles, California 90095, USA.}

\begin{abstract}

{\color{black}The Dawn spacecraft recently observed the presence of dark material, which in turn proved to be associated with the presence of OH and H-rich material, on the surface of Vesta.} The source of this dark material has been almost unanimously identified with the low albedo asteroids, likely analogous to the carbonaceous chondrites found on Earth, that impacted on Vesta over its lifetime. However, it is still a matter of debate whether the delivery of the dark material is associated with a few {\color{black}large impact events, to micrometeorites or to the continuous, secular flux of impactors on Vesta}. The ``continuous flux'' scenario, in particular, predicts that a {\color{black}significant} fraction of the exogenous material accreted by Vesta {\color{black}should} be due to non-dark impactors likely analogous to ordinary chondrites, which {\color{black}instead represent only a minor contaminant} in the Howardite-Eucrite-Diogenite meteorites. In this work, we explored the ``continuous flux'' scenario and its implications for the composition of the vestan regolith{\color{black}, taking advantage of the data from the Dawn mission and the Howardite-Eucrite-Diogenite meteorites to constrain the contamination history of Vesta}. We developed a model for the delivery of the exogenous material to Vesta and {\color{black}verified} how the results it supplies are sensitive to the different parameters we consider. {\color{black}We calibrated the flux of impactors predicted by our model with the number of dark craters observed inside the Rheasilvia basin and we tested the assumptions on the impact conditions by studying the formation of Cornelia crater and of its dark deposits with a hydrocode simulation. We used our calibrated model to show that the ``stochastic events'' scenario and the ``micrometeoritic flux'' scenario are just natural consequences of the ``continuous flux'' scenario.} We then used the model to estimate the amounts of dark and hydroxylate materials that were delivered on Vesta since the Late Heavy Bombardment and we showed how our results match well with the values estimated by the Dawn mission. We finally used our model to assess the amount of Fe and siderophile elements that the continuous flux of impactors would mix in the vestan regolith{\color{black}: concerning the siderophile elements, we focused our attention on the role of Ni.} The results we obtained are in agreement with the data available on the Fe and Ni content of the Howardite-Eucrite-Diogenite meteorites and {\color{black} can be used as a reference frame in future studies of the data from the Dawn mission and of the Howardite-Eucrite-Diogenite meteorites. Our model cannot yet provide an answer to the conundrum of the fate of the missing non-carbonaceous contaminants, but we discuss some possible reasons for this discrepancy with the otherwise coherent picture described by our results.}
%More generally, our results provide a clearer global picture of the role of impacts in the processes of mass exchange and contamination in the asteroid belt.

\end{abstract}

\begin{keyword}
Asteroid Vesta \sep Impact processes \sep Asteroids, surfaces
\sep Regoliths \sep Meteorites
\end{keyword}

\end{frontmatter}

% \linenumbers

\section{Introduction}
\label{intro}

{\color{black}The Dawn spacecraft recently observed the presence of dark material on the surface of Vesta \citep{mccord2012,reddy2012} and the dark material proved, in turn, to be associated with H-rich material \citep{prettyman2012} and OH \citep{desanctis2012}.} While the source of this dark material has been almost unanimously identified with the carbonaceous chondrites, particularly the CM and CR chondrites that have been observed as clasts inside the Howardite-Eucrite-Diogenite (HED) family of meteorites (see \citealt{mccord2012,reddy2012,prettyman2012,desanctis2012} for more in-depth discussions), the actual delivery scenario is still debated.

\citet{mccord2012} linked the delivery of the dark material to the flux of low albedo impactors associated with the collisional history of Vesta since the Late Heavy Bombardment. \citet{reddy2012} also associated the delivery of the dark material to impacting asteroids{\color{black}: however, instead of a continuous flux,} these authors proposed that the delivery {\color{black}could be} due to a few  {\color{black}large, low-velocity impact events (at least two)}, one of these being {\color{black}responsible for} the formation of the Veneneia basin. As an alternative possibility, \citet{reddy2012} proposed that the dark material could be delivered by micrometeorites. {\color{black}As these authors pointed out, however, the micrometeoritic flux since the Late Heavy Bombardment is too low to account for the observed amount of dark material on Vesta.} As the flux of dark micrometeorites was likely orders of magnitude more intense during the Late Heavy Bombardment, \citet{reddy2012} suggested that the ancient meteoritic flux could have significantly contributed to the total budget of the dark material together with the impacts of low albedo asteroids. 

In discussing the detection of OH in the spectral features of Vesta, \citet{desanctis2012} linked its presence in the vestan regolith either to low-velocity impactors or to the micrometeoritic flux. These authors argued against a more or less continuous flux of OH-carrying impactors and favoured instead a temporally limited delivery, possibly located in the more ancient past of Vesta \citep{desanctis2012}. In discussing the distribution and setting of H-rich materials and studies of carbonaceous chondrite clasts in howardites, \citet{prettyman2012} pointed out instead that the H content in the vestan regolith plausibly rules out a single, isolated impact or a  temporally limited enhancement of the meteoritic flux as possible sources. These authors argued that the concentration and distribution of H {\color{black}suggests on the contrary} an accumulation over time from numerous impactors and asteroidal dust \citep{prettyman2012}.

In this work we will focus on the role of asteroidal impacts on Vesta in delivering the dark material, and the OH and H-rich material detected by the Dawn spacecraft. {\color{black}The ``stochastic events'' and the ``micrometeoritic flux'' scenarios discussed by \citet{reddy2012} and \citet{desanctis2012}  are not necessarily in contrast with the ``continuous flux'' scenario discussed by \citet{mccord2012}. As already noted by \citet{mccord2012}, in the case of a continuous flux of impactors about half of the dark material would be delivered by a handful of large asteroids and an even larger contribution would be associated with stochastic events like the one responsible for Veneneia basin. Moreover, the ``continuous flux'' scenario naturally incorporates the ``micrometeoritic flux'' scenario.} The ``continuous flux'' scenario, however, predicts that the dominant fraction of the exogenous material accreted by Vesta would be due to non-dark impactors (see e.g. \citealt{mccord2012}), likely analogous to ordinary chondrites.  {\color{black}These contaminants have been detected only in marginal quantities in the HED family of meteorites \citep{lorenz2007}, raising the conundrum of their fate.}

{\color{black}The aim of this work is to address the problem of the contamination history of Vesta by providing a quantitative assessment of the amounts of the different exogenous materials delivered in the ``continuous flux'' scenario. %across the collisional history of the asteroid. % and comparing them with all observational constraints provided by the Dawn mission. 
To achieve this goal, we improved the model we first used in \citet{mccord2012} and refined the calculations performed there to estimate the flux of impactors and the amount of dark material delivered on Vesta since the Late Heavy Bombardment. We tested the assumptions on the average impact velocity and angle by studying the formation of Cornelia crater with a hydrocode simulation and verifying that the distribution of the dark material inside the crater is satisfactorily reproduced by the remnants of the impactor material in the simulation}. We extended our physical model to allow the assessment of the amounts of OH and H-rich material and of non-dark exogenous material delivered to Vesta. {\color{black}Concerning the latter class of contaminants,} we focused on the role of Fe and of Ni, which we used as our tracer for the siderophile elements based on the results of \citet{warren2009}. We finally compared our results with the findings of the Dawn mission \citep{desanctis2012,desanctis2012b,mccord2012,prettyman2012,reddy2012,yamashita2013} and discussed their implications for the composition of the vestan regolith in light of our current understanding of the HED  family of meteorites (e.g. \citealt{zolensky1996,lorenz2007,warren2009}).

\begin{table}
\begin{center}
\begin{tabular}{cc}
\hline
\bf{D$_{i}$ (in km)} & \bf{N$_{i}$} \T\B\\
\hline
$    0.98  $     &     $   394278.90  $   \T\\
$    1.23  $     &     $   296503.10  $     \\
$    1.55  $     &     $   221080.10  $     \\
$    1.96  $     &     $   162026.40  $     \\
$    2.46  $     &     $   115400.30  $     \\
$    3.10  $     &     $    78939.80  $     \\
$    3.91  $     &     $    51398.50  $     \\
$    4.92  $     &     $    31739.60  $     \\
$    6.19  $     &     $    18630.70  $     \\
$    7.79  $     &     $    10463.90  $     \\
$    9.81  $     &     $     5671.80  $     \\
$   12.40  $     &     $     2992.30  $     \\
$   15.60  $     &     $     1548.00  $     \\
$   19.60  $     &     $      789.70  $     \\
$   24.60  $     &     $      554.00  $     \\
$   31.00  $     &     $      338.00  $     \\
$   39.10  $     &     $      224.00  $     \\
$   49.20  $     &     $      185.00  $     \\
$   61.90  $     &     $      164.00  $     \\
$   77.90  $     &     $      116.00  $     \\
$   98.10  $     &     $       91.00  $     \\
$  123.50  $     &     $       64.00  $     \\
$  155.50  $     &     $       38.00  $     \\
$  195.70  $     &     $       17.00  $     \\
$  246.40  $     &     $        8.00  $     \\
$\geq300.00$     &     $        6.00  $   \B\\
%$  310.20  $     &     $        3.00  $     \\
%$  390.50  $     &     $        1.00  $     \\
%$  491.60  $     &     $        2.00  $     \\
%$\geq618.90$     &     $        1.00  $   \B\\
\hline
\end{tabular}
\caption{Present-day population of the asteroid belt. Column $1$ is the central diameter estimated by \citet{bottke2005a} for each bin from the magnitudes reported by \citet{jedicke2002}, while column $2$ is the incremental number of asteroids in each bin from \citet{jedicke2002}. The total population of asteroids larger than $\sim 1$ km is therefore estimated to be $N_{belt}\approx1.39 \times 10^{6}$. Table adapted from \citet{bottke2005a}. {\color{black}Note that, differently from \citet{bottke2005a}, the bins associated with diameters larger than $\sim300$ km have been grouped into a single class as their populations are dominated by small number statistics}.}\label{sfd}
\end{center}
\end{table} 

\section{Method}

{\color{black} As we mentioned previously, our delivery scenario is based on the one we first employed in \citet{mccord2012}. As such, it uses the intrinsic impact probability of Vesta together with a description of the temporal evolution of the population of the asteroid belt to statistically assess the number and sizes of the impactors on the asteroid. Using these values together with scaling laws for the retention efficiency of Vesta, it is then possible to estimate the amount of exogenous material accreted by the asteroid.

Following \citet{mccord2012}, we used our current understanding of the present-day fraction of dark and non-dark asteroids to constrain the amount of potential carriers of dark material among the projectiles hitting Vesta. However, with respect to \citet{mccord2012} we also used the available information on the composition of the different classes of meteorites to try to quantitatively assess the amounts of the different materials delivered to the asteroid that could be measured by the Dawn mission.

In the following sections we will describe in detail all the aspects of our model. Before proceeding, we must point out that some of the observational parameters (e.g. the fraction of dark asteroids) and of the theoretical results (e.g. the mass fraction of the impactors that remains on Vesta) we used to build our model are still poorly constrained. As a consequence, in the model we considered different possibilities for these parameters in order to assess how much our results are affected by these uncertainties.}

\subsection{Temporal evolution of the asteroid population and flux of impactors on Vesta}\label{methods-population}

In their work, \citet{mccord2012} analytically estimated the flux of dark impactors on Vesta based on the present day population of bodies with $D \geq 1$ km in the asteroid belt (see Table \ref{sfd} and \citealt{bottke2005a}). 

The authors took into account the depletion in the population of asteroids that should have occurred across the last $3.5$ Ga due to chaotic diffusion (estimated to be of a factor $2$, \citealt{minton2010}) by assuming a constant decay (i.e. a linear decrease) in the number of asteroids over time. From the point of view of the total number of impacts (but not from that of their temporal distribution), this is equivalent to assuming a population constant in time that is $1.5$ times larger than the present one. Using this approximation, \citet{mccord2012} estimated the flux of impactors on Vesta over the last $3.5$ Ga by multiplying the number of impacts per unit time $F_{i}$ for the integration time.% $\Delta T=3.5$ Ga. 

Following \citet{obrien2011}, if we consider $N_{i}$ impactors of diameter $D_{i}$ (see Table \ref{sfd}) the impact frequency is $F_{i} = P_{V} A_{i} N_{i}$, where $P_{V} = 2.72\times10^{-18}$ km$^{-2}$ yr$^{-1}$ is the intrinsic impact probability of Vesta \citep{obrien2011}, $A_{i}= (R_{V} + 0.5D_{i})^{2}$ is the cross-sectional area of Vesta and the impactors (the factor $\pi$ is included into $P_{V}$, \citealt{bottke1994}) and $R_{V}=262.7$ km is the mean radius of Vesta \citep{russell2012}. 

As a consequence, \citet{mccord2012} estimated the number of impacts $n_i$ due to $N_{i}$ impactors of a given size as 
\begin{equation}\label{mccord}
n_{i}=1.5\ F_{i}\ \Delta T = 1.5\ P_{V} A_{i} N_{i} \ \Delta T
\end{equation}
where $\Delta T = 3.5\times10^{9}$ years {\color{black} and the factor $1.5$ is included to take into account the depletion of the asteroid belt, as discussed above.}

In this work, we adopted a different approach, which more correctly accounts for the temporal evolution of the population of asteroids over time. Instead of considering the population as constant, we took advantage of the exponential decay law $f_{t}=C(t/ 1\,year)^{-D}$ derived by \citet{minton2010}, where $f_{t}$ is the fraction of surviving bodies after the time $t$ expressed in years while $C=1.9556$ and $D=0.0834$ are constants\footnote{{\color{black}According to \citet{minton2010}, the piecewise logarithmic law they derived is more accurate than the power law here used. The coefficients reported for the logarithmic law in their paper, however, are incorrect: the value of coefficient B$_{2}$ should be $0.03468\pm0.00005$ (D. Minton, pers. comm.).}}. Note that this equation or, more properly, these coefficients are valid (i.e. they reproduce the depletion rate of the asteroid belt) for $1$ Ma $<t<3.98$ Ga \citep{minton2010}. Under these constraints, the value of $f$ varies between $0.618$ ($t=1$ Ma) and $0.309$ ($t=3.98$ Ga, i.e. now).

Using these values and the present-day population of asteroids $N_{i}$ from Table \ref{sfd}, we can extrapolate a primordial population of asteroids $N_{ip}$ so that at each given time $t$ we have $N_{it}=f_{t}N_{ip}$. This approach is made possible by the results of \citet{bottke2005a,bottke2005b}, who showed that the present SFD of the asteroid belt should be stable over the temporal intervals we are considering. {\color{black}Note that, as in \citet{mccord2012}, in this work we also focused our attention on the asteroids with estimated diameter $D\gtrapprox 1$ km, as shown in Table \ref{sfd}. The reasons for this choice is threefold.

First, we chose to base our model on the catalogued asteroid population and on the data from the Sloan Digital Sky Survey (SDSS) as reported  by \citet{jedicke2002}. Therefore, while we take advantage of the magnitude to diameter conversion made by \citet{bottke2005a}, we are restricting ourselves only to the observational data and we are not including the results of modelling efforts (see e.g. \citealt{bottke2005b}). 

Second, the number of sub-km asteroids is still poorly constrained. \citet{gladman2009} report that the differential slope of the smaller asteroids, as estimated through the Sub-Kilometer Asteroid
Diameter Survey (SKADS), is possibly steeper than previously thought (see also \citealt{obrien2011}). Specifically, \citet{gladman2009} suggest that this uncertainty in the differential slope could result in an uncertainty of a factor $2-3$ in the real number of sub-km asteroids.

Third, notwithstanding this uncertainty \citet{gladman2009} report that the differential slope should be of the order of $-2.5$. Values of the differential slope higher than $-3$ imply that the population of asteroids does not increase fast enough, when moving toward smaller sizes, to compensate for the decrease in mass due to the lower diameters of the bodies (e.g. \citealt{davis1979}). Consequently, the mass contribution of sub-km asteroids will be more limited than that of km-sized asteroids.} 

{\color{black}The number of impacts $n_i$ due to an evolving population of $N_{it}$ impactors of a given size can therefore be computed as
\begin{equation*}
n_{i} =  \int P_{V} A_{i} N_{ip} f_{t} dt
\end{equation*}
In this work, we evaluated the previous integral over the desired temporal interval $\Delta T$ as a sum over discrete timesteps $\Delta t = 1000$ years, so that the previous equation becomes
\begin{equation}\label{population}
n_{i} = \sum_{j} P_{V} A_{i} N_{ip} f_{t} (t_{0}+ j\Delta t)
\end{equation}
where $t_{0}=1$ Ma and $0 \leq j \leq \Delta T/\Delta t$.}

{\color{black}In the following, we will consider two temporal intervals $\Delta T$: the post-Late Heavy Bombardment period or, more properly, the last $3.98$ Ga, and the post-Rheasilvia period, i.e. the last $1$ Ga \citep{schenk2012,marchi2012}. This choice is motivated by the following reasons. 

First, following the time the average impact velocity on Vesta reached its present value, impacts removed from the asteroid several times more material than they brought (see e.g. \citealt{svetsov2011}). \citet{turrini2013}, based on the results of \citet{mccord2012} and of S. Pirani (Master Thesis at the University of Rome ``La Sapienza''), pointed out that the combined effects of the impacts on Vesta during the Late Heavy Bombardment (as estimated in the scenario discussed by \citealt{minton2009}) and during the following $4$ Ga could saturate the surface of the asteroid to a level equal to or slightly larger than the one currently observed in the oldest terrain ($\sim10\%$, \citealt{marchi2012}). In this case, most exogenous material deposited before the Late Heavy Bombardment could have been removed by later impacts.

Second, the collisional evolution of Vesta and, in general, of the asteroid belt before $\sim 4$ Ga ago is still debated.  \citet{obrien2007} showed that the dynamical friction between planetesimals and planetary embryos could result in a depletion rate of the asteroid belt lower than that considered in the calculations of \citet{bottke2005b}, to now the most complete model of the collisional evolution of the asteroid belt. Also, the implications of the Late Heavy Bombardment in the framework of the updated, self-consistent version of the Nice Model \citep{levison2011} and those of a possible now-extinct extended population of the asteroid belt \citep{bottke2012} for the collisional history of Vesta have not been assessed to date. Dedicated studies would therefore be required before it is possible to reliably extend the model to earlier times.

Finally, the delivery of the exogenous material to Vesta at the time of the formation of its crust represents a separate chapter. Vesta \citep{turrini2011,turrini2013} and, more generally, the asteroid belt \citep{turrini2012} underwent a phase of enhanced collisional evolution at the time of the formation of Jupiter. {\color{black}\citet{turrini2011} and \citet{turrini2013} showed that the flux of impactors on Vesta could amount, in mass, to up to $\sim$$10\%$ %$$
 of the present mass of the asteroid.} However, as discussed by \citet{turrini2013}, at that time Vesta still possessed a limited solid crust overlying a mostly molten interior (see e.g. \citealt{formisano2013}), so the fate of these contaminants is still to be assessed. Moreover, the collisional evolution of Vesta during the phase of depletion of the asteroid belt (see \citealt{coradini2011} and \citealt{obrien2011} and reference therein) plausibly removed most if not all of the exogenous material previously deposited on the surface.

These reasons are at the basis of our decision to focus on the post-Late Heavy Bombardment temporal interval, for which the collisional evolution of the asteroid belt is better constrained (see e.g. \citealt{coradini2011} and \citealt{obrien2011} and references therein). As the Dawn mission provided a constraint on the age of the Rheasilvia basin ($1$ Ga, \citealt{schenk2012,marchi2012}) and the excavation of this basin most likely removed all previous exogenous contaminants (see \citealt{schenk2012}, \citealt{jutzi2013} and \citealt{ivanov2013}), thus providing us with a ``clean slate'' from the point of view of the contamination, we decided to consider the last $1$ Ga as an additional case.}

{\color{black}
\subsection{Characterization of the impacts and of the different kind of impactors}\label{method-impacts}

As we are interested in the global effects of the collisional history of Vesta for the contamination of its surface, in our model we assumed that all impacts take place at the average impact velocity $V_{i}=4.75$ km/s estimated by \citet{obrien2011} and occur at $45\degree$ respect to the local normal to the surface of the asteroid \citep{melosh1989}. For the impactors, we adopted average densities of $1400$ kg/m$^{3}$ for dark asteroids (i.e. the average density of C type asteroids from \citealt{britt2002,carry2012}), of $2700$ kg/m$^{3}$ for non-dark asteroids (i.e. the average density of S type asteroids from \citealt{britt2002,carry2012}) and of $2400$ kg/m$^{3}$ when considering all possible impactors (based on our reference fraction of dark asteroids in the asteroid belt described in Sect. \ref{method-dark_and_water}).

Note that these densities are significantly lower (a few $10\%$) than the bulk densities of carbonaceous and ordinary chondrites \citep{consolmagno2008,macke2011a}. In particular, the density of dark impactors is $\sim30\%$ lower than the average bulk density of the CM-CR-CI meteorites ($\sim2100-2200$ kg/m$^{3}$, see \citealt{macke2011a} and Sect. \ref{method-dark_and_water}) that are the potential carriers of water to Vesta, as we will discuss in Sect. \ref{method-dark_and_water}. Nevertheless, the values we adopted will provide us with a conservative estimate of the amounts of exogenous materials delivered to Vesta over the temporal intervals here considered.
}

\subsection{Mass retention efficiency}\label{method-retention}

In their work, \citet{mccord2012} evaluated the fraction of the mass of the impacting bodies that is retained by Vesta by using Eq. $8$ (valid for impact velocities $V \leq 15$ km/s) from \citet{svetsov2011},  i.e.
\begin{equation}\label{retention_svetsov}
f_{r} = \left(0.14 + 0.003V\right)\ln v_{esc} + 0.9V^{-0.24}
\end{equation}
where $v_{esc}=0.37$ km/s is the escape velocity from Vesta \citep{turrini2011} and $V=4.75$ km/s is the average impact velocity of the impactors \citep{obrien2011}. %{\color{black} Note that Eq. $8$ from \citet{svetsov2011} provided the escaped fraction of the projectile's mass, i.e. $1-f_{r}$. 
{\color{black}\citet{mccord2012} estimated that about $46.6\%$ of the impacting mass would be retained by Vesta ($f_{r}=0.466$). Therefore, if  $m_{i}$ is the mass of the impacting body, the mass retained by Vesta is thus $m_{r} = f_{r}\ m_{i}=0.466\ m_{i}$.} Note that integrating the retention efficiency over the range of possible impact velocities computed by \citet{obrien2011} gives similar results to using solely the average impact speed, the difference in the retention factor being of the order of $10\%$.

{\color{black}In their study of the delivery of volatile materials to the Moon, \citet{ong2010} estimated an average retention efficiency of about $16.5\%$ for asteroidal impactors. This retention efficiency was evaluated assuming a median impact velocity on the Moon of $20$ km/s and averaging over all possible impact velocities.} {\color{black}It must be noted that the study of \citet{ong2010} focused mainly on cometary impactors, for which they provided the retention efficiencies at different impact velocities, and considered also the contribution of the re-condensation of water vapors. For asteroidal impactors, they instead provided only the average value previously reported. As the behavior of cometary impactors and asteroidal impactors is not necessarily comparable and for Vesta it is unlikely that the re-condensation of vapors played a significant role in the contamination of the surface of the asteroid (see e.g. \citealt{turrini2011}), we focused only on the single value of the average retention efficiency reported for asteroidal impactors.} To estimate the difference between the results of \citet{ong2010} and those of \citet{svetsov2011}, we computed the average retained mass fraction for the Moon using the scaling laws reported by \citet{svetsov2011}. Following \citet{ong2010}, we used an escape velocity of $2.38$ km/s and an average impact velocity of $20$ km/s. The logarithmic interpolation of Eqs. $8$ and $9$ from \citet{svetsov2011} for the chosen impact velocity gives an average retained mass of $55.7\%$. Therefore, the scaling laws by \citet{svetsov2011} result in a retention efficiency about $3.4$ times larger than the results of \citet{ong2010}.

{\color{black}Before proceeding, it must be stressed that linearly scaling Eq. \ref{retention_svetsov} to the results of \citet{ong2010} obtained for an average impact velocity of $20$ km/s is not necessarily correct over the range of impact velocities characteristic on Vesta ($1-10$ km/s, \citealt{obrien2011}). However, as we are mainly interested in understanding how much the uncertainty on the retention scaling law can affect the results of the model, in the following we will also consider the following case:
\begin{equation}\label{retention_ong}
\begin{split}
f^{*}_{r} & = 0.296\ f_{r} \\
& = 0.296 \times (\left(0.14 + 0.003V\right) \\
& \quad \times \ln v_{esc} + 0.9V^{-0.24})
\end{split}
\end{equation}
Using the same escape velocity and average impact speed we used for Eq. \ref{retention_svetsov}, Eq. \ref{retention_ong} gives a retention efficiency $f^{*}_{r}=0.138$.} {\color{black} The retention efficiencies supplied by Eqs. \ref{retention_svetsov} and \ref{retention_ong} will be used to transform the fluxes of impactors (computed as described in Sect. \ref{methods-population}) into mass fluxes on Vesta. Using the abundances of the different kind of impactors and their compositions, which we will describe in Sects. \ref{method-dark_and_water} and \ref{method-siderophiles}, we will then use to the mass fluxes thus obtained to estimate the contamination of the surface of Vesta by different classes of exogenous materials.}

\subsection{Dark impactors, OH and H-rich material}\label{method-dark_and_water}

\citet{mccord2012} estimated the fraction of low albedo impactors on Vesta computing the ratio between asteroids with B, C, P and D spectral types and all spectroscopically classified asteroids from the JPL Small-Body Database Search Engine\footnote{\url{http://ssd.jpl.nasa.gov/sbdb_query.cgi}}. The value they obtained was {\color{black}$F_{d}=0.22$}. An independent estimate performed by \citet{desanctis2012} consistently gave $0.1 \leq F_{d} \leq 0.3$.

In this work, we used the catalogue of asteroids from the JPL Small-Body Database Search Engine to assess the fraction of low albedo impactors on Vesta in three different ways, i.e.: $1$) using the same method as \citet{mccord2012}; $2$) using the same method as \citet{mccord2012} but including only asteroids with C, D and P spectral types; $3$) computing the fraction of asteroids with albedo lower than or equal to $0.05$ among all asteroids for which a value of albedo has been estimated. As mentioned above, method $1$ held a value of {\color{black}$F_{d1}=0.22$}. Method $2$ held a value of {\color{black}$F_{d2}=0.18$}. Method $3$ held a value of {\color{black}$F_{d3}=0.28$. The number of dark impactors of a given size $D_{i}$ over a temporal interval $\Delta T$ on Vesta is then
\begin{equation}\label{dark_num}
n_{i}^{dark}=n_{i} \times F_{d}
\end{equation}
where $n_{i}$ is obtained though Eq. \ref{population} and $F_{d}$ is one of the three values previously discussed. The amount of dark material delivered by these impactors can instead be expressed as
\begin{equation}\label{dark_mass}
m_{i}^{dark}=\left( \frac{\pi}{6} D_{i}^{3} \rho \right) \times f_{r} \times n_{i}^{dark}
\end{equation}
where $f_{r}$ is obtained either through Eq. \ref{retention_svetsov} or Eq. \ref{retention_ong} and $\rho$ is the average density of the dark impactors.} In the following we will use the value derived by \citet{mccord2012} and through our method $1$ as our reference value. We will use the results of methods $2$ and $3$ to discuss {\color{black}how the uncertainty on the abundance of dark asteroids in the asteroids belt affects the results of the model}.

{\color{black}Coincident with that of the dark material, the} Dawn mission reported the presence of OH \citep{desanctis2012} and H-rich material \citep{prettyman2012} on Vesta. The results of \citet{prettyman2012} and \citet{desanctis2012} indicated a clear correlation between the presence of dark material and that of OH and H-rich material. {\color{black}Both H and OH are likely present in the vestan regolith in the form of hydrated minerals and not as water. However, in meteoritics (see e.g. \citealt{robert2003}) and in nuclear spectroscopy (see e.g. \citealt{prettyman2007}) their presence is generally quantified in terms of the equivalent amount of water needed to reproduce the measurements. In the following, therefore, we will refer to these two (likely overlapping) classes of materials with the common term of ``water-equivalent material.''}

{\color{black}As discussed by \citet{mccord2012}, the dark impactors are plausibly asteroids similar in composition to the carbonaceous chondrites.}
The most efficient carriers of water-equivalent material among {\color{black}the dark impactors should therefore be those bodies similar in composition to the CM, CR and CI carbonaceous chondrites \citep{jarosewich1990,robert2003}. In particular,} CM chondrites and, to a smaller extent, CR chondrites represent the dominant components of carbonaceous chondritic clasts observed in HED meteorites \citep{zolensky1996,lorenz2007}. The CR/CM ratio in HED meteorites, moreover, is the same as the one observed among modern CM and CR falls on Earth \citep{zolensky1996}. Therefore, in order to assess the amount of water-equivalent material that a continuous flux of impactors would bring on Vesta, we tried to estimate the frequency of CM-, CR- and CI-like projectiles among {\color{black}the dark impactors} hitting the asteroid. Based on the observation by \citet{zolensky1996} that the CR/CM ratios are similar among meteorite falls on Earth and HEDs meteorites, we used the fluxes of the falls for the different {\color{black}classes of} carbonaceous chondrites on Earth as our planetary analogue. From the Meteoritical Bullettin\footnote{\url{http://www.lpi.usra.edu/meteor}}, as of May 2013, we note that CM chondrites represent {\color{black}$34\%$} of the falls of carbonaceous chondrites, CR chondrites represent {\color{black}$6.8\%$} and CI chondrites {\color{black}$11.4\%$}. {\color{black}The fraction of the masses of these impactors that represent water-equivalent material can be assumed, on average, $R_{w}=0.1$, i.e. $10$ wt$\%$ \citep{jarosewich1990,robert2003}.}

If, following \citet{zolensky1996}, we focus only on CM and CR chondrites, we obtain that a fraction {\color{black}$F_{w}^{CM+CR}=0.41$} of the dark impactors on Vesta would bring water-equivalent material to the asteroid. This will be our standard case. Should we include also CI chondrites, i.e. we assume that our database of carbonaceous clasts in HED meteorites is not complete, the fraction of carriers of water-equivalent material would rise to {\color{black}$F_{w}^{CM+CR+CI}=0.53$}. We caution the readers, however, that while this is possibly the best approximation we can make to date, it is not necessarily a good or physically meaningful one. As an example, CI chondrites are very fragile and the bulk of them likely do not survive the passage through the Earth's atmosphere, causing the underestimation of their contribution. Moreover, there is no guarantee that the modern flux of meteorites on Earth is a meaningful approximation of the secular flux in the asteroid belt over one or more Ga{\color{black}, as it can be significantly affected by temporally-limited events like the break-up of an asteroid.}

{\color{black}In order to better constrain the implications of the uncertainty on the flux of carriers of water-equivalent material for our results, we considered also} the water delivery efficiency discussed by \citet{ong2010}. In their work, these authors assumed that about one-third of impacting asteroids would be C-type {\color{black}(our dark impactors, i.e. $F_{d}=0.33$)} and that two-thirds of those impactors would contain water-equivalent material {\color{black}($F_{w}=0.67)$}. These authors assumed an average $10$ wt$\%$ water content, i.e. the same value we adopted in the previous cases. As these authors point out, {\color{black}their assumptions result in water representing} $2.2\%$ of {\color{black}the total mass flux due to impacts}. As a comparison, in our reference case we have $F_{d1}\times F_{w}^{CM+CR} \times R_{w}=0.9\%$, i.e. our {\color{black}reference} water delivery efficiency is $2.5$ times smaller.

{\color{black}Using the value of $R_{w}$ and one of the possible values of $F_{w}$ previously discussed, the amount of water-equivalent material $m_{i}^{w}$ delivered on Vesta by dark impactors of a given size $D_{i}$ over a temporal interval $\Delta T$ is then
\begin{equation}\label{water_mass}
m_{i}^{w}=m_{i}^{dark} \times F_{w} \times R_{w}
\end{equation}
where $m_{i}^{dark}$ is obtained through Eq. \ref{dark_mass}}.
%there are $1583$ known samples of carbonaceous chondrites, of which $460$ are CM and $9$ are CI. CM chondrites thus represent $29.06\%$ of all known carbonaceous chondrites, while CI only represent $0.57\%$. We therefore assumed in our model that a fraction $F_{w}=0.2963$ of dark impactors bring water and hydrated material to the asteroid and that the fraction of their masses that represent water-equivalent material is $R_{w}=0.1$, i.e. $10\%$ \citep{jarosewich1990,robert2003}.

\subsection{Fe and siderophile elements}\label{method-siderophiles}

As the Gamma Ray and Neutron Detector (GRaND) on-board the Dawn spacecraft measured the global Fe content of minerals on the surface of Vesta \citep{prettyman2012,yamashita2013}, we used our model to assess what a continuous flux of impactors would imply from the point of view of the contamination of the vestan regolith by Fe and siderophile elements. Concerning the latter, we focused our analysis on the role of Ni, which has been suggested by \citet{warren2009} to play an important role as tracer of the regolith's maturity and enrichment.

{\color{black}In contrast to the cases of the dark material and the water-equivalent material discussed in Sect. \ref{method-dark_and_water}, the howarditic material composing most of the vestan regolith \citep{desanctis2012b,prettyman2012} would contain  amounts of Fe and Ni, even if uncontaminated. The mean Fe contents for diogenites and basaltic eucrites are respectively $13$ wt$\%$ and $14.7$ wt$\%$ (\citealt{jarosewich1990}; \citealt{prettyman2012}, Supplementary Material). A $2$:$1$ eucrite-diogenite mixture, suggested to be the average compositional mixture of howardites \citep{warren2009}, would have an average Fe content of $14.1$ wt$\%$, which is consistent with the mean Fe content of $13.8$ wt$\%$ of howardites (\citealt{jarosewich1990}; \citealt{prettyman2012}, Supplementary Material). In order to be able to compare our results with the measurements of GRaND, we will follow \citet{yamashita2013} and assume that the native Fe content of the vestan regolith is $13.8$ wt$\%$.}

{\color{black}The Ni content of the different materials composing the vestan crust spans a greater range of values than Fe. The Ni content of diogenites varies between $5-200$ $\mu$g/g, i.e. $5\times10^{-4}-0.02$ wt$\%$ \citep{warren2009}. Monomict and cumulate eucrites have a lower Ni content, ranging between $0.1-10$ $\mu$g/g \citep{warren2009}, i.e. $10^{-5}-10^{-3}$ wt$\%$. Polymict eucrites have a Ni content similar to that of diogenites, with one exceptional polylmict eucrite sample as rich in Ni as $\sim1000$ $\mu$g/g, i.e. $0.1$ wt$\%$ \citep{warren2009}. Howardites are characterized by a large variability in Ni content and overlap the range of diogenites and polymict eucrites. According to \citet{warren2009}, Ni content in howardites varies in the range $10-5000$ $\mu$g/g, i.e. $10^{-3}-0.5$ wt$\%$. If we assume, based on the values reported by \citet{warren2009}, an average Ni content of $1$ $\mu$g/g for eucrites and $40$ $\mu$g/g for diogenites, the resulting $2$:$1$ mixture will have a Ni content of $\sim14$ $\mu$g/g. This is also the order of magnitude of the Ni content of the more Ni-poor samples of howardites analyzed by \citet{warren2009} and will be used in the following as our native Ni content of the vestan crust.

Contrary to HED meteorites, the order of magnitude of the Fe and Ni contents of chondrites (both ordinary and carbonaceous) is much less variable. Based on the values reported by \citet{jarosewich1990}, the average Fe and Ni contents of all (both dark and non-dark) impactors in our model can be assumed to be respectively $20$ wt$\%$ (i.e. $F_{Fe}=0.2$) and $1$ wt$\%$ (i.e. $F_{Ni}=0.01$). The amount of Fe delivered on Vesta by impactors of a given size $D_{i}$ over a temporal interval $\Delta T$ can then be expressed as
\begin{equation}\label{Fe_mass}
m_{i}^{Fe}= \frac{m_{i}^{dark}}{F_{d}} \times F_{Fe}
\end{equation}
where $m_{i}^{dark}$ is obtained through Eq. \ref{dark_mass} and is divided by the value $F_{d}$ to obtain the total mass delivered by dark and non-dark impactors. A similar equation holds for the case of Ni.}

{\color{black}
\subsection{Crater saturation and ejecta blanketing}\label{method-saturation}

The effects of crater saturation and of the blanketing due to crater ejecta are two important factors that must be considered in discussing the contamination of Vesta. %order to have a clearer picture of the survival of the exogenous contaminants on the surface of the asteroid and of the possibility to observe them with the instruments on-board the Dawn spacecraft. 
{\color{black}The estimates of the extent of the vestan surface affected by craters and of the level of crater saturation reached over a given timespan allow to assess how long} the dark material can survive on the surface of Vesta before being removed by the impacts of non-dark asteroids and whether we should expect to be able to link most dark material to specific cratering events or not. The assessment of the degree to which crater ejecta cover the surface of the asteroid, on the other hand, would supply information on how much of the dark material is going to be buried to depths that cannot be probed by the instruments on-board the Dawn spacecraft and how old the dark material on the surface can be.  

Given that our model already computes the flux of impactors hitting Vesta over the desired timespan in order to estimate the degree of contamination produced,   we complemented it with a collisional model similar to the one described in \citet{turrini2013}, to estimate the craters produced by the impacts and the degree of saturation they cause. Moreover, we included in the model also an estimate of the ejecta blanketing associated with the craters and of the surface area affected by the fragments of the dark impactors along the line of what we already did in \citet{mccord2012}, as described in their Supplementary Information.  

The diameter of the craters produced by the flux of impactors was estimated using the following scaling law for rocky targets by \citet{holsapple2007}:
\begin{align}\label{crater_law}
\frac{R_{c}}{r_{i}}=0.93\left(\frac{g\,r_{i}}{V^{2}}\right)^{-0.22}	\left(\frac{\rho_{i}}{\rho_{v}}\right)^{0.31}\nonumber\\
+0.93\left(\frac{Y_{v}}{\rho_{v} V^2}\right)^{-0.275}\left(\frac{\rho_{i}}{\rho_{v}}\right)^{0.4}
\end{align}
where $R_{c}$ is the final radius of the crater, $r_{i}$ is the radius of the impactor, $g=0.22$ m\,s$^{-2}$ is the surface gravity of Vesta \citep{reddy2012}, $V=3.36$ km/s is the average vertical impact velocity (to take into account the fact that impacts are assumed to take place at $45\degree$ respect to the normal to the surface, see Sect. \ref{method-impacts}), $Y_{v}=7.6$ MPa is the strength of the material composing the surface of Vesta (assumed to behave as soft rock, \citealt{holsapple1993}), $\rho_{i}$ is the density of the impactor and $\rho_{v}=3090$ kg\,m$^{3}$  is the density of the crust of Vesta \citep{russell2012,russell2013}.

%Using the diameters of the craters estimated this way, we computed the associated R-value distribution. Following the \citet{crater1978}, we assumed an initial crater diameter $D_{1}=1$ km and we considered bins where $D_{i+1}=\sqrt{2}D_{i}$. For each $[D_{i},D_{i+1})$ bin we computed the number of craters $N_{i}$ and the geometric mean diameter 
%\begin{equation}
%\overline{D_i}=\left( \prod_{j=1}^{N_{i}} d_{j} \right)^{\frac{1}{N_{i}}} \nonumber
%\end{equation}
%where $d_{j}$ are the diameters of the individual craters. Then the R-value associated with each bin is computed as
%\begin{equation}\label{rvalue}
%R_{i}=\frac{N_{i}\,\overline{D}^{3}}{S_{v}\left(D_{i+1}-D_{i}\right)}
%\end{equation}
%In building the R-value distributions, we considered only those bins where the cumulative number of craters was greater than or equal to $1$.

As we mentioned previously, using the diameters computed with Eq. \ref{crater_law}, we also computed the cumulative surface affected by ejecta blanketing and the darkened surface due to the fragments of the dark impactors.
To compute the overall blanketed surface we followed \citet{melosh1989} and assumed that, on average, ejecta cover a surface with radius twice as large as that of the associated crater. The total blanketed surface $A_{B}$ then becomes
\begin{equation}\label{blanketed}
A_{B}=\sum_{i}\pi \left( D_{i}^{2}-\left(0.5D_{i}\right)^{2} \right)=\sum_{i} \frac{3\pi}{4} D_{i}^{2}
\end{equation}
where $D_{i}$ are the diameters of the individual craters
and we summed over the craters produced by all impactors (as also dark impactors excavate ejecta that can cover previously deposited dark material) during the relevant temporal interval, and we subtracted the contribution of the craters themselves, as in principle they can expose previously buried dark material. 

To compute the darkened surface, we followed instead \citet{mccord2012} and assumed that the fragments from the dark impactors would distribute inside the crater and inside a cone $30\degree$ wide and extending up to $4$ times the radius of the crater downstream to the impact direction. As a consequence, the total darkened surface is
\begin{align}\label{darkened}
A_{B} & =\sum_{i} \left( \frac{\pi}{4} D_{i}^{2} + \frac{\pi}{48}\left( \left(4D_{i}\right)^{2}-D_{i}\right)^{2} \right) \nonumber\\
      & =\sum_{i} \left( \frac{\pi}{4} D_{i}^{2} + \frac{5\pi}{16}D_{i}^{2} \right)=\sum_{i} \frac{9\pi}{16} D_{i}^{2}
\end{align}
where we sum only over the craters produced by dark impactors.
}

\subsection{Cratering erosion}\label{method-erosion}

{\color{black}The assessment of the mass loss of Vesta from the formation of its basaltic crust to now is important to understand the global picture of the erosion of the surface of the asteroid (see e.g. \citealt{mcsween2013} for a discussion concerning Rheasilvia and \citealt{turrini2013} for a more general one) and can play a significant role in unveiling the history of the asteroid belt and the Solar System \citep{coradini2011,turrini2011,turrini2012,turrini2013}. 

Understanding the balance between mass gain and mass loss due to impacts {\color{black}(see e.g. \citealt{svetsov2011} for an updated discussion in different regimes of impact and escape velocities)} is also fundamental to constrain how long the exogenous material can survive on the surface of Vesta and how ancient the dark deposits and veneers observed by the Dawn mission are. As a consequence, we used our model to assess the mass loss due to the cumulative flux of impactors over the two temporal intervals considered in this work.%last $3.98$ Ga and $1$ Ga.%, to complement the study of the primordial erosion of Vesta started by \citet{turrini2013}.}

The constraints on the thickness of the vestan regolith are still limited and the available estimates range from about $100$ m to about $1$ km (\citealt{jaumann2012}, see also their Supplementary Materials). Using Eq. \ref{crater_law} for the different size bins reported in Table \ref{sfd} and multiplying the results for the average depth-to-diameter ratio ($0.168$) estimated for Vesta by \citet{vincent2013}, we can see that all but possibly the smaller impactors here considered would excavate significantly deeper than the regolith layer. As a consequence, in our model we used the scaling law for rock from \citet{holsapple2007} to estimate the mass loss of Vesta.

%{\color{black}Following \citet{svetsov2011} we averaged the scaling law of mass loss for cohesive soil from \citet{holsapple2007} over all possible impact angles and we obtained
%\begin{equation}\label{erosion}
%f_{e}=\frac{M_{e}}{m_{i}}=0.01\left(\frac{V}{v_{esc}}\right)^{1.23}\left(\frac{\rho_{i}}{\rho_{reg}}\right)^{0.2}=0.25
%\end{equation}
%where $M_{e}$ is the eroded mass, $m_i$ is the mass of the impactor, $V=4.75$ km/s is the average impact velocity, $v_{esc}=0.37$ km/s is the escape velocity from Vesta, $\rho_i=2400$ kg/m$^{3}$ is the average bulk density assued for all impactors and $\rho_{reg}=2000$ kg/m$^{3}$ is the assumed density of the regolith layer on Vesta.}

%As a comparison, using a fixed impact angle of $45^{\circ}$ in the scaling law from \citet{holsapple2007} instead of averaging over all possible impact angles gives similar results, with values about $20\%$ larger. 
{\color{black}From this scaling law in the  angle-averaged form computed by \citet{svetsov2011} we obtain 
\begin{equation}\label{erosion}
f_{e}=\frac{M_{e}}{m_{i}}=0.03\left(\frac{V}{v_{esc}}\right)^{1.65}\left(\frac{\rho_{i}}{\rho_{V}}\right)^{0.2}=2.13
\end{equation}
where $M_{e}$ is the eroded mass, $m_i$ is the mass of the impactor, $V$ is the average impact velocity (see Sect. \ref{method-impacts}), $v_{esc}=0.37$ km/s is the escape velocity from Vesta \citep{turrini2011}, $\rho_i=2400$ kg/m$^{3}$ is the average density assumed for all impactors (see Sect. \ref{method-impacts}) and $\rho_{reg}=3090$ kg/m$^{3}$ is the assumed density of the crust (regolith plus bedrock) of Vesta \citep{russell2012,russell2013}.

The total eroded mass due to impactors of a given size $D_{i}$ over a temporal interval $\Delta T$ can then be expressed as
\begin{equation}
m_{i}^{erod.}=\left( \frac{\pi}{6} D_{i}^{3} \rho_{i} \right) \times n_{i} \times f_{e}
\end{equation}
where $n_{i}$ is obtained from Eq. \ref{population} and $f_{e}$ is obtained from Eq. \ref{erosion}. We can already see, however, that cratering erosion will be between about $5$ times (if Eq. \ref{retention_svetsov} is used for the retention efficiency) and $15$ times (if Eq. \ref{retention_ong} is used instead) more efficient than mass accretion {\color{black} for a body the mass of Vesta and the characteristic impact velocities of the asteroid belt (see also Fig. 3 of \citealt{svetsov2011}).}
%While we will focus on Eq. \ref{erosion} in our analysis, the interested readers may convert the collisional erosion from one regime to the other simply by multiplying our results by a factor $9.7768$.

{\color{black}
\subsection{Hydrocode modeling and validation of the impact parameters}\label{methods-hydrocode}

To verify whether the average impact velocity and impact angle we assumed for all impactors are appropriate choices to study the delivery of the dark material, we complemented our model with a hydrocode simulation devised to test the validity of our impact parameters. 
%Numerical simulations allow us to study large scale impact cratering events, which are not possible to reproduce in laboratory. 
To perform the simulation we took advantage of the iSALE-3D shock physics code \citep{elbeshausen2009,elbeshausen2011}, based on the solver described in \citet{hirt1974}. The code includes a strength model \citep{collins2004,melosh1992,ivanov1997} and a porosity compaction model \citep{wunnemann2006,collins2011} and its development history is described in \citet{elbeshausen2009}.

We used iSALE-3D to simulate the formation of Cornelia crater (Lat: 9.4~S, Lon: 225.6~E), one of the dark craters identified and studied by \citet{reddy2012}. In our simulation the mesh is composed by $400$ cells in the horizontal direction, $200$ cells in the vertical one and $100$ cells in depth, with a  spatial resolution of $100$~m (including both the impactor and the target layer with some additional space to allow the motion of the excavated material). The impact scenario assumes a spherical dunite impactor with diameter of $1.6$~km, porosity of $30\%$ and a final density of $2300$ kg/m$^{3}$ (consistent with a CM-like impactor, \citealt{macke2011a}) hitting Vesta at $4.75$~km/s with an impact angle of $45\degree$. The {\color{black}vestan surface is represented} by a basalt layer of about $15$~km having density of $2650$ kg/m$^{3}$ (consistent with a howarditic or eucritic layer with about $15\%$ porosity, \citealt{consolmagno2008}) and surface gravity of $0.25$~m/s$^2$. The choice of the composition of the projectile is limited to those materials existing in the simulation package and having proper equation of state. However, as we showed, the density and the porosity of the projectile have been adjusted to better fit a carbonaceous chondrite.
}

\section{Results}

\begin{table*}[t]
\begin{center}
\begin{tabular}{cccc}
\hline
%{} & \multicolumn{3}{c}{\bf{N. of Impacts}} \T\B\\
%\hline
%\bf{Impactors} &
\head{Temporal} & \head{Evolving} & \head{Linear Decay} & \head{{\color{black}Steady-State}}    \T \\
%{} & 
\head{Interval} & \head{Asteroid Belt} & \head{\citep{mccord2012}} & \head{Asteroid Belt} \B \\
\hline
\multicolumn{4}{c}{\head{All Impactors}}\T\B\\
%All  & 
$3.98$ Ga & $1141.9$ & $1383.3$ & $1046.4$ \T \\
%All  &
$1.00$ Ga & $265.3$  & $300.4$  & $262.0$ \B \\
\multicolumn{4}{c}{\head{Dark Impactors}}\T\B\\
%Dark & 
$3.98$ Ga & $244.4$ & $297.1$ & $223.0$  \T \\
%Dark & 
$1.00$ Ga & $55.5$  & $64.5$  & $54.8$  \B \\
\hline
\end{tabular}
\caption{Number of impacts on Vesta due to all impactors and to dark impactors only over the last $3.98$ Ga and the last $1$ Ga. The column ``Evolving Asteroid Belt'' reports the values obtained assuming a decaying population in the asteroid belt as in \citet{minton2010}, while the column ``{\color{black}Steady-State} Asteroid Belt'' reports those obtained assuming a constant population equal to the present one. The column ``Linear Decay'' shows the corresponding values obtained using the linear decay approximation adopted by \citet{mccord2012}.}\label{impacts}
\end{center}
\end{table*}

{\color{black}As discussed in Sect. \ref{methods-population}, we run our model over two temporal intervals: the post-Late Heavy Bombardment period, i.e. the last $3.98$ Ga, and the post-Rheasilvia period, i.e. the last $1$ Ga. Before using the model to assess the contamination of Vesta (Sects. \ref{results-scenarios}--\ref{results-bright_contaminants}), however, we estimated the uncertainties that characterize our results (i.e. the impact and mass fluxes,  Sects. \ref{results-fluxes} and \ref{results-uncertainty}) and validated the model and its assumptions against the observational features of the dark material on the vestan surface (Sects. \ref{results-rheasilvia} and \ref{results-cornelia}).}

{\color{black}
\subsection{The temporal evolution of the asteroid belt and Vesta's collisional history}\label{results-fluxes}

\begin{figure}
\centering
+\includegraphics[width=\columnwidth]{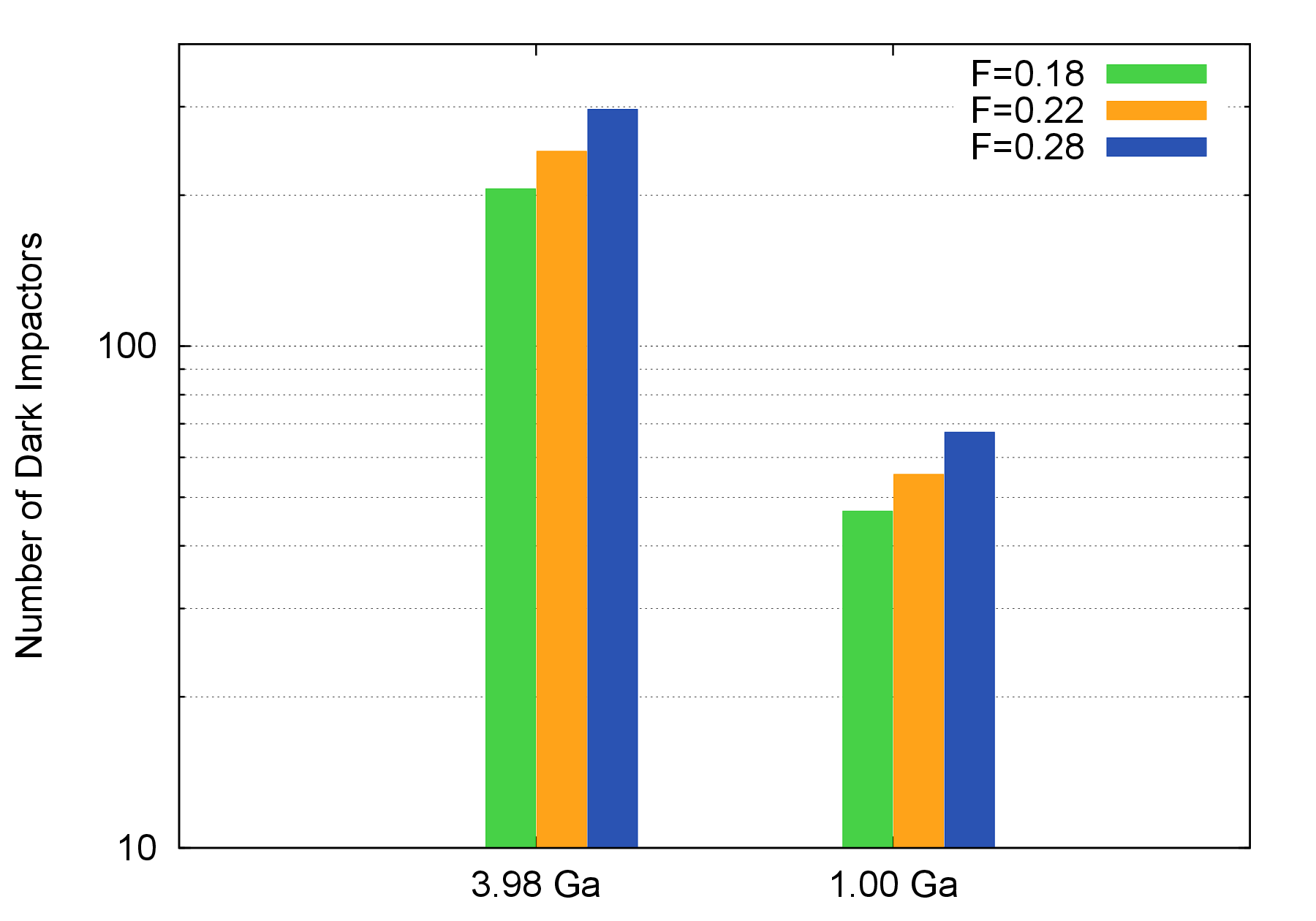}
\caption{Number of dark impactors on Vesta over the last $3.98$ Ga and the last $1$ Ga for the different values of the fraction of dark asteroids $F_{d}$ in the asteroid belt we described in Sect. \ref{method-dark_and_water}.}\label{dark-range}%{fig-masses}
\end{figure}

\begin{figure*}[t]
\centering
\includegraphics[width=\textwidth]{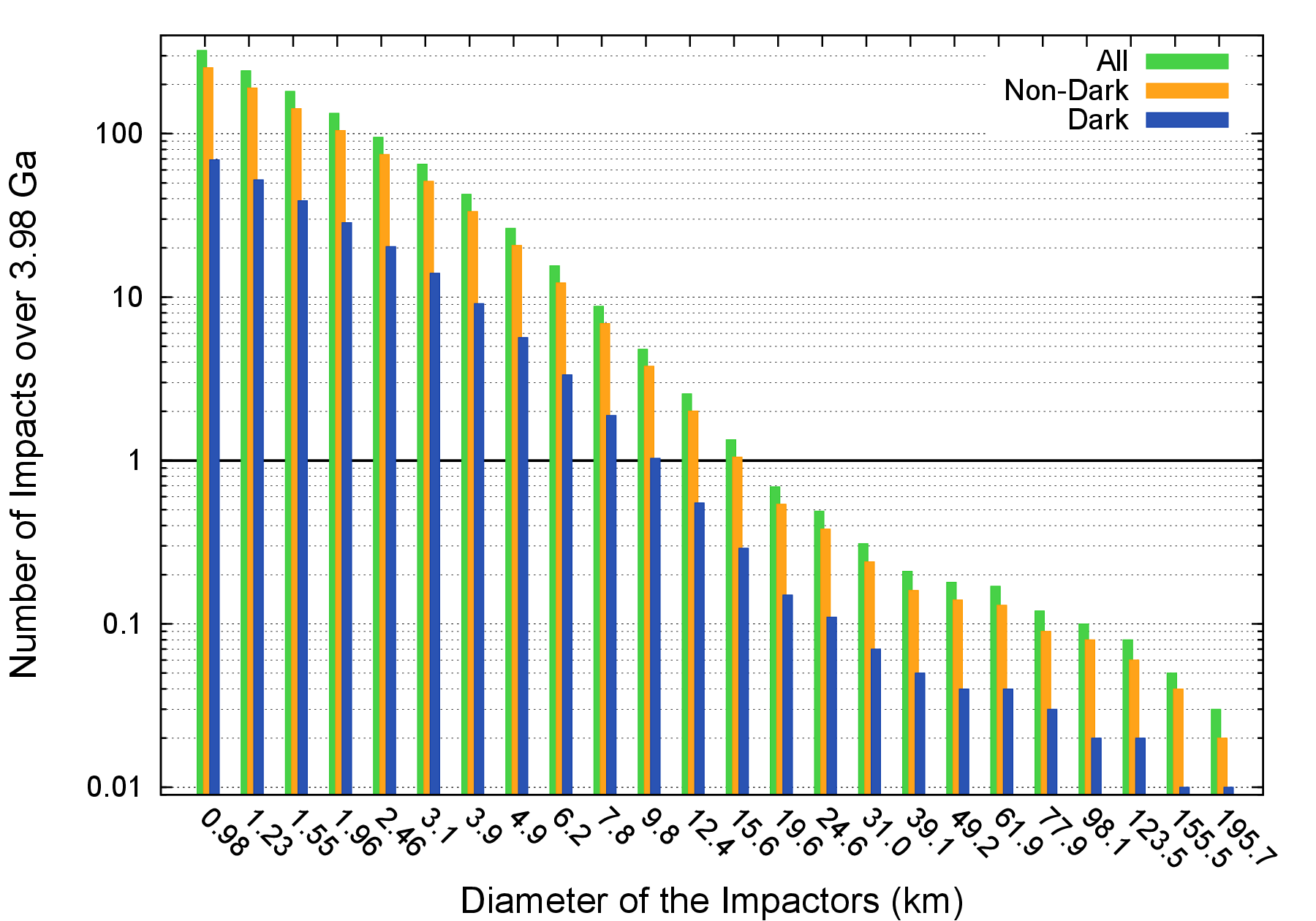}
\caption{Number of impacts over $3.98$ Ga for dark impactors, non-dark impactors and for all impactors for the different size classes from Table \ref{sfd}. We do not account for size classes associated with a number of impacts lower than $1$ (i.e. stochastic events) in the assessments of the contamination (see Figure \ref{mflux}).}\label{nflux}
\end{figure*}

The first comparison we performed was with the number of impacts estimated by \citet{mccord2012} and the number of impacts that would take place on Vesta if the population of the asteroid belt was {\color{black}in a steady state} at its present level. The results we obtained are summarized in Table \ref{impacts}. As can be seen by comparing the different columns in Table \ref{impacts}, the assumption of a {\color{black}steady-state} asteroid belt produces results that are closer to those obtained considering a more realistic, exponentially decaying asteroid belt than those obtained assuming a linearly decaying population of asteroids, as instead we did in \citet{mccord2012}. The {\color{black}steady-state} asteroid belt gives a number of impacts that is $10\%$ lower than our reference case of an exponentially decaying asteroid belt, while the linearly decaying asteroid belt results in a number of impacts that is $20\%$ larger than the latter. 

We then verified how much the total number of dark impactors is affected by our assumption on the fraction of dark asteroids in the asteroid belt. In Figure \ref{dark-range} we show the number of dark impactors expected on Vesta over the two temporal intervals here considered using the different values for the fraction of dark asteroids we discussed in Sect. \ref{method-dark_and_water}. As can be seen, the number of dark impactors changes by $\pm20\%$, but the order of magnitude of the results is the same. As a consequence, in the following we will always implicitly refer to our reference case where $F_{d}=0.22$.

\begin{figure*}[t]
\centering
\includegraphics[width=\textwidth]{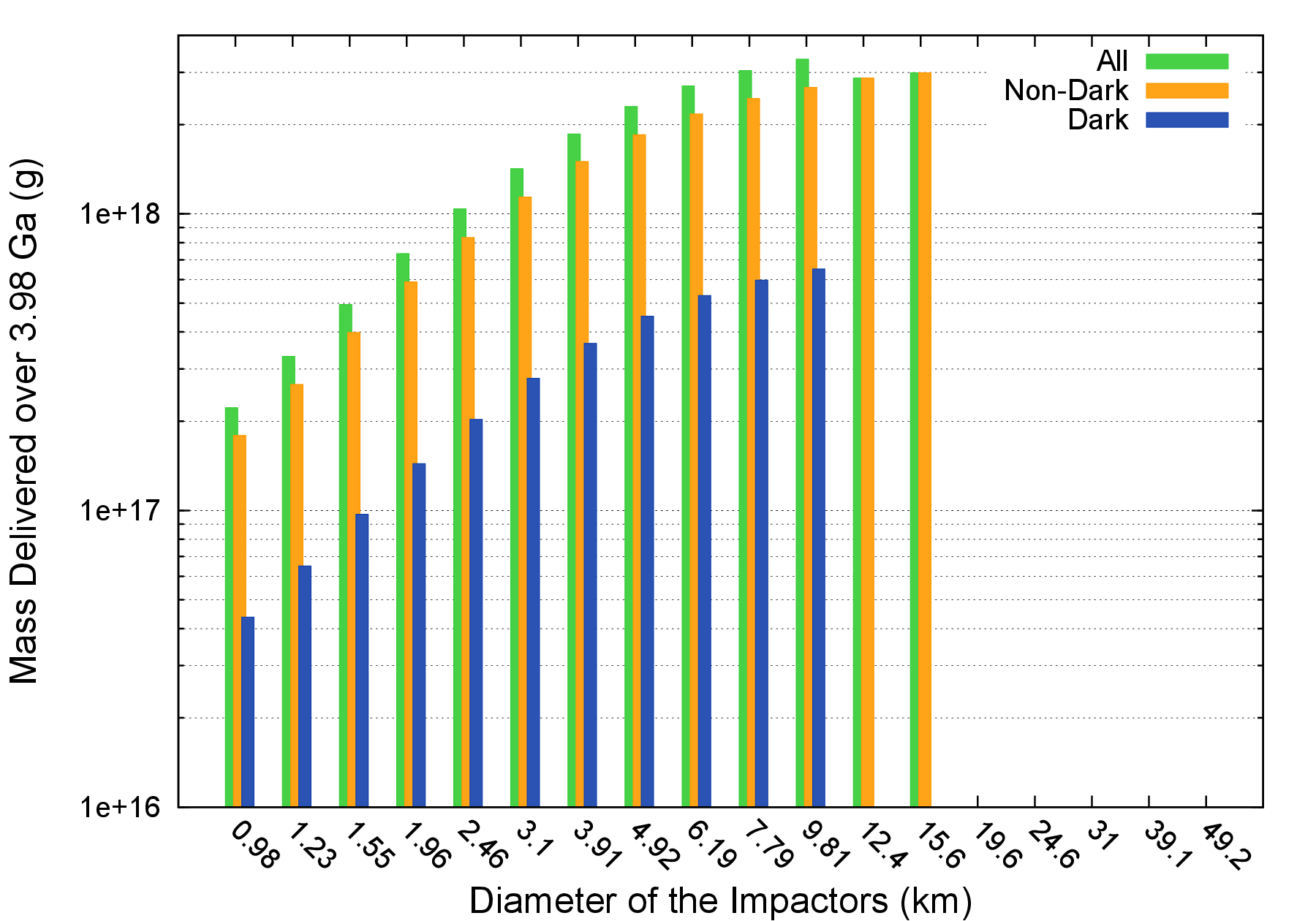}
\caption{Masses (in g) delivered to Vesta over $3.98$ Ga by dark impactors, non-dark impactors and by all impactors. The mass retention efficiency is computed using Eq. \ref{retention_svetsov} so it is the reference retention efficiency we considered in this work. As mentioned in the caption of Figure \ref{nflux}, the delivered masses are computed only for those size classes of impactors that produce at least $1$ impact over the considered temporal interval.}\label{mflux}%{fig-masses}
\end{figure*}

In Figs. \ref{nflux} and \ref{mflux} we show respectively the number of impacts and the mass delivered by dark, non-dark and by all possible impactors as computed using Eq. \ref{retention_svetsov} for the different size bins from Table \ref{sfd}. In Figure \ref{nflux} we show only those size bins that are associated with at least a $1\%$ chance of producing one impact over the considered temporal interval. In Figure \ref{mflux} we show instead only those size bins that are associated with at least $1$ impact over the considered timespan. 

As over the last $1$ Ga the exponentially decaying asteroid belt and the {\color{black}steady-state} one produce essentially the same results within $1\%$, in Figs. \ref{nflux} and \ref{mflux} we show only the values referring to the last $3.98$ Ga. Interested readers can obtain the number of impacts and the mass fluxes of dark and non-dark impactors over the last $1$ Ga simply by using the values of Table \ref{sfd} together with Eq. \ref{mccord} (where the factor $1.5$ should be dropped as the population of asteroids should be considered {\color{black}in a steady-state}) and Eqs. \ref{dark_num} and \ref{dark_mass}.
}

{\color{black}
\subsection{Uncertainties on the number of impacts and on the delivered masses}\label{results-uncertainty}

The uncertainties associated with our estimates of the fluxes of impactors on Vesta are governed by Poisson statistics, therefore $1\sigma=\sqrt{N}$ where $N$ is the population of impactors under consideration. Over the last $3.98$ Ga, the $1\sigma$ uncertainty on the flux of all possible impactors is of about $3\%$, while in the case of the dark impactors the $1\sigma$ uncertainty is of about $6\%$. Because of the lower number of events, the $1\sigma$ uncertainties on the fluxes of all impactors and of dark impactors over the last $1$ Ga become respectively about $6\%$ and about $13\%$.  

However, a comparison of Figs. \ref{nflux} and \ref{mflux} immediately reveals that the delivery of the dark material and, more generally, of all exogenous materials is dominated by the contribution of the largest impactors. In particular, the largest impactors for both dark and non-dark asteroids are associated with a single impact event over the relevant timespan (see e.g. Figure \ref{nflux}). According to Poisson statistics, the uncertainty on these impact events is of the same order as the number of events themselves, i.e. the events may or may not have taken place over the considered temporal interval. The uncertainty on the single largest dark impactor in Figure \ref{nflux} translates in a $30\%$ uncertainty on the total budget of the dark material delivered over the last $3.98$ Ga (see Figure \ref{mflux}). Analogously, the uncertainty on the single largest non-dark impactor in Figure \ref{nflux} implies a $15\%$ uncertainty on the total budget of the exogenous (i.e. dark and non-dark) material delivered over the same timespan (see Figure \ref{mflux}). Over the last $1$ Ga, the uncertainties in the number of the largest impactors translate in uncertainties of about $20\%$ on the total delivered masses of dark and non-dark materials. 

{\color{black}The uncertainty on the contribution of the largest impactors is also affected by our choice of considering only those size bins from \citet{bottke2005a} that produce at least $1$ impact over the considered timespans. As can be seen from Fig. \ref{nflux} and as we pointed out in the Supplementary Information of \citet{mccord2012}, dark asteroids with diameters comprised between $10$ km and $25$ km have $10-60\%$ probabilities of hitting Vesta over the last $3.98$ Ga and can cumulatively produce one impact on the asteroid. If we average their contribution to the total mass of dark material delivered to Vesta by weighting over their impact probability, we obtain they could bring about the same amount of dark material as all dark asteroids smaller than $10$ km cumulatively (\citealt{mccord2012}, Supplementary Information). This is not surprising, as the differential slope of the asteroidal size-frequency distribution implies that the population of asteroids in each size bin does not decrease fast enough to compensate for the increase in mass of the asteroids when we move toward larger diameters (see e.g. Fig. \ref{mflux}).

Given that this behaviour is determined by the size-frequency distribution of the asteroids, the same issue applies when we consider the contamination due to all possible impactors (i.e. dark and non-dark). Asteroids with diameter comprised between $20$ km  and $60$ km can produce, cumulatively, two impacts on Vesta over the last $3.98$ Ga and can deliver about three times the amount of exogenous material globally delivered by all impactors smaller than $20$ km. It is interesting to note that these two possible large impacts would be consistent with the kind of events proposed to have caused the formation of Veneneia and Rheasilvia \citep{marzari1996,ivanov2013,jutzi2013}. 

As a consequence, even when we consider only a specific set of values for the parameters of the model (e.g. $F_{d}$, $f_r$, etc.), these uncertainties on the numbers of impactors imply that the $1\sigma$ errors affecting the delivered amounts of exogenous materials cannot be smaller than $20-30\%$ and can grow to a factor of a few. Removing or limiting the effects of these uncertainties will require the calibration of the collisional model used in this work with the complete crater population of Vesta and the use of dedicated hydrocode simulations to assess the retained mass for the largest impacts, as the scaling laws we are using in our model do not necessarily hold for these events.  When comparing the predictions of our model to the data supplied by the Dawn mission or by the laboratory studies of the HED meteorites, therefore, we will limit ourselves to consider only the orders of magnitude of the relevant quantities.
}

{\color{black}
\subsection{Dark craters on Vesta and inside Rheasilvia}\label{results-rheasilvia}

The Dark Material Units catalogue (Palomba et al., 2013, this issue) produced by analysing the data supplied by VIR \citep{desanctis2011,desanctis2012b} has revealed that more than $50\%$ of the occurrences of the dark material are associated with impact-related features (crater rims, crater slopes, crater walls, ejecta). If we look to the global distribution of the dark craters and of the regions characterized by the lowest reflectance values we show in Figure \ref{reflectance-map} (where we considered the reflectance at 1.4 $\mu$m as our proxy), we can see that both the highest density of dark craters and lowest reflectance values are associated with the oldest, highly-cratered terrains of Vesta \citep{marchi2012}. 

This is in agreement with the distribution of the H-rich material mapped by GRaND \citep{prettyman2012}, with the distribution of the OH mapped by VIR \citep{desanctis2012} and with the scenario of secular accumulation of the dark material due to a continuous flux of dark impactors. The older (from the cratering point of view) a terrain is, the more dark material will accumulate since the last resetting event. More recent terrains instead had their content of exogenous material reset and had less time to accumulate it again. The ages of the oldest terrains, however, are still not well constrained \citep{marchi2012}; as a consequence, we cannot use them as a reliable test for our model from the point of view of the expected number of dark features. %(things are different when we consider the delivered amount of dark material, as we will discuss later). 

%It is also interesting that the fraction of the surface of Vesta characterized by the highest presence of H-rich material ($\sim30\%$, \citealt{prettyman2012}), which broadly coincides with the fraction of the vestan surface characterized by values of the reflectance at 1.4 $\mu$m lower than $0.28$ ($\sim27\%$, see Figure \ref{reflectance-map}), is consistent with the order of magnitude of the darkened fraction of the vestan surface ($\sim27\%$) estimated by \citet{mccord2012} for the period going from the Late Heavy Bombardment to now.

The age of the Rheasilvia basin, however, is much better constrained ($1$ Ga, \citealt{marchi2012,schenk2012}). The Rheasilvia basin, moreover, represents the optimal test-bed for this kind of comparison, as the extensive excavation \citep{jutzi2013} caused by its formation and the formation of the partially underlying Veneneia basin erased all pre-existing deposits of exogenous material. This basin is therefore a ``clean slate'' that recorded all impact events due to dark impactors over the last $1$ Ga. The first test against the observational data from the Dawn mission that we performed on our model, therefore, was to compare the number of dark impactors hitting Vesta in the last $1$ Ga with the number of dark craters observed inside the Rheasilvia basin.  

\begin{figure*}[t]
\includegraphics[width=\textwidth]{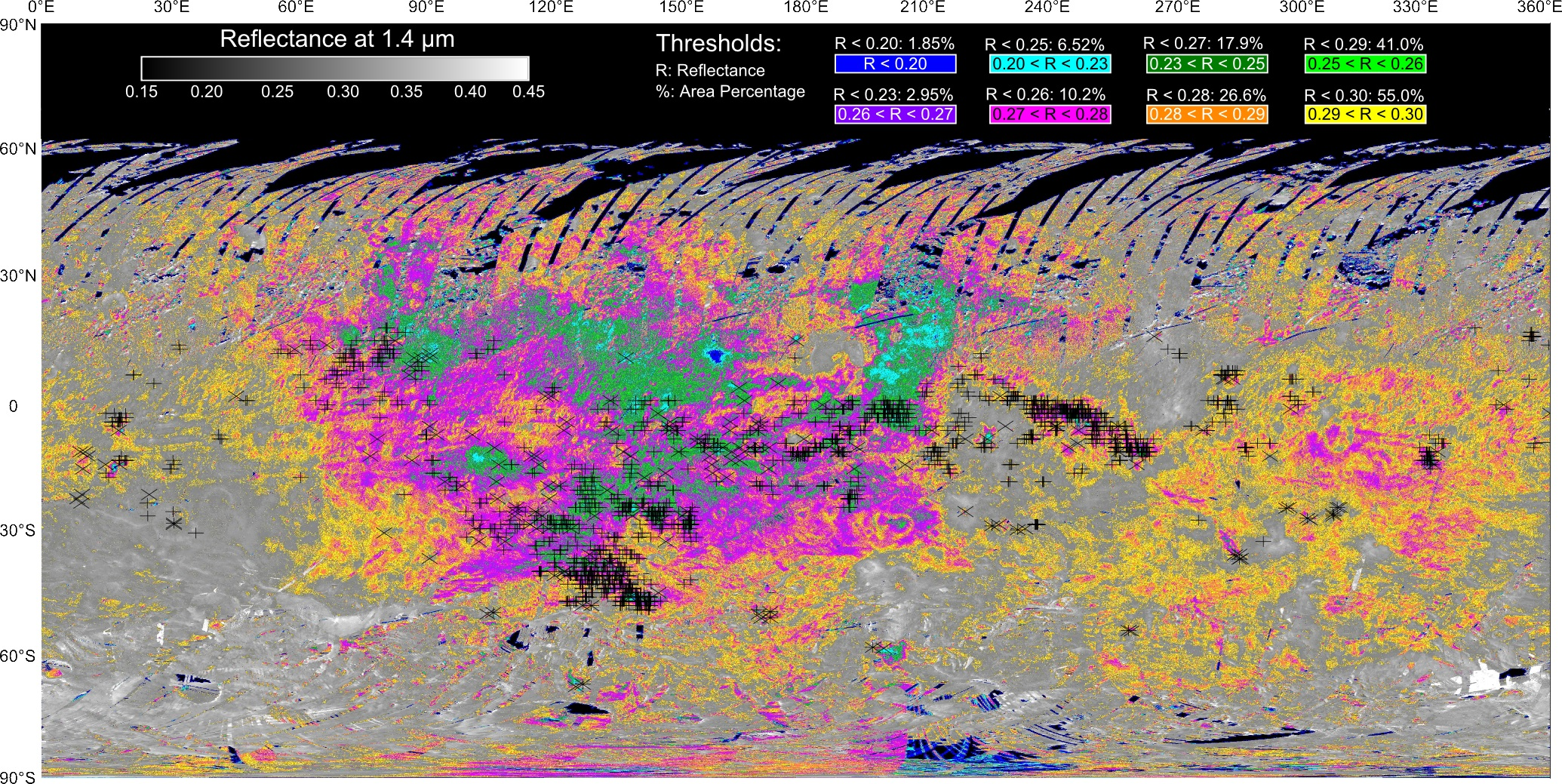}
\caption{{\color{black}Distribution of the dark material on the surface of Vesta as revealed by the visual and near-infrared spectrometer VIR on-board the Dawn spacecraft using the spectral reflectance at $1.4$ $\mu$m as our proxy. Black symbols indicate the dark craters and the dark spots identified with the color data from  the Framing Camera on-board the Dawn spacecraft during the HAMO phase at about 60 m/pixel \citep{reddy2012}.} Also indicated are the area percentages of the surface of Vesta that are associated with the different threshold levels of reflectance we selected.}\label{reflectance-map}
\end{figure*}

The Rheasilvia basin has a diameter of $\sim500$ km \citep{schenk2012}: a simple back-of-the-envelope calculation shows that its surface accounts for $22.6\%$ of the total surface of Vesta computed from its mean radius $R_{V}=262.7$ km. This implies that, statistically, $12.5\pm3.5$ dark impactors should fall inside Rheasilvia in our reference case for the flux of dark projectiles.
If we consider all the different values of $F_{d}$ that we discussed in Sect. \ref{method-dark_and_water} and Sect. \ref{results-fluxes}, we obtain a minimum of $10.6\pm2$ and a maximum of $15.3\pm4$ of dark craters inside Rheasilvia, i.e. a variability of about $20\%$ respect to our reference value. A comparison with Figs. $11$ and $12$ from \citet{reddy2012} reveals $12$ dark craters inside or overlapping the rim of Rheasilvia. If we consider only the dark craters that \citet{reddy2012} identified inside the Rheasilvia basin, the total drops to $10$ dark craters. Finally, an updated count that we performed in the framework of this study resulted in $16$ dark craters inside or on the rim of the Rheasilvia basin. All these values are consistent (generally within $1\sigma$) with the range of possible outcomes resulting from our model, supporting the scenario of a continuous and isotropic flux of dark impactors on Vesta and our choice of parameters.

As a side product of this test of the reliability of our model, we can take advantage of the impact fluxes we computed to estimate the probability of the scenario assumed by \citet{jutzi2013} for the formation of Veneneia and Rheasilvia. \citet{jutzi2013} assumed that the two basins formed due to the impact of $\sim60$ km wide asteroids, colliding with Vesta with velocities slightly above the average impact velocity computed by \citet{obrien2011}. The probability of Vesta being hit by an asteroid with this diameter over the last $3.98$ Ga is $\sim16\%$. The chances of Vesta being hit by two such asteroids in the same temporal interval, therefore, are of the order of $2-3\%$. If we require that one of the two impactors is a low albedo asteroid, as proposed by \citet{reddy2012}, the chances of this scenario can drop down to $0.5\%$. It must be noted, however, that \citet{reddy2012} assumed that Veneneia was formed by a smaller impactor than the one of \citet{jutzi2013}.

{\color{black}
\subsection{Cornelia crater as a case study and a calibration test}\label{results-cornelia}

In their work, \citet{reddy2012} used the Cornelia crater as a case study to illustrate the spectral behavior and the distribution of the dark material associated with a crater (see Figure $7$ in \citealt{reddy2012}). To investigate whether the dark material inside Cornelia could have been delivered by a low albedo impactor hitting Vesta with the average impact velocity and angle we assumed in our model, we simulated the formation of Cornelia using the iSALE-3D shock physics code as detailed in Sect. \ref{methods-hydrocode}. 

The result of our simulation is presented in Figure \ref{fig-cornelia}. The crater  resulting from the simulation has a depth of $2.5$~km and a diameter of about $11$~km, which is comparable to the depth of $3.7$~km and the diameter of $12.6$~km measured for Cornelia by \citet{vincent2013}. As can be seen in Figure~\ref{fig-cornelia}, at the end of the simulation the distribution of the projectile's material inside the crater reproduces reasonably well the bilobate distribution of the dark material, the region devoid of dark material upstream of the crater center and their symmetry respect to the impact direction. The main differences between the results of the simulation and the dark material inside Cornelia are due to a series of subsequent landslides along the crater walls, easily identifiable as a set of mostly circular, brighter regions among the dark material near the top of the crater walls.

A small fraction of the projectile's material falls outside the rim of the crater downstream of the impact direction, covering an area of about $11.52$ km$^{2}$. This is less than what observed by \citet{reddy2012}, but this difference can be due either to our choice of the physical properties of target and impactor or to the fact that in the specific case of Cornelia the impactor had a higher velocity or a higher (respect to the local vertical) impact angle. The match between the real and the simulated Cornelia craters is nevertheless sufficiently good to confirm our choice of the impact parameters in the model.
}

\begin{figure*}[t]
\includegraphics[width=\textwidth]{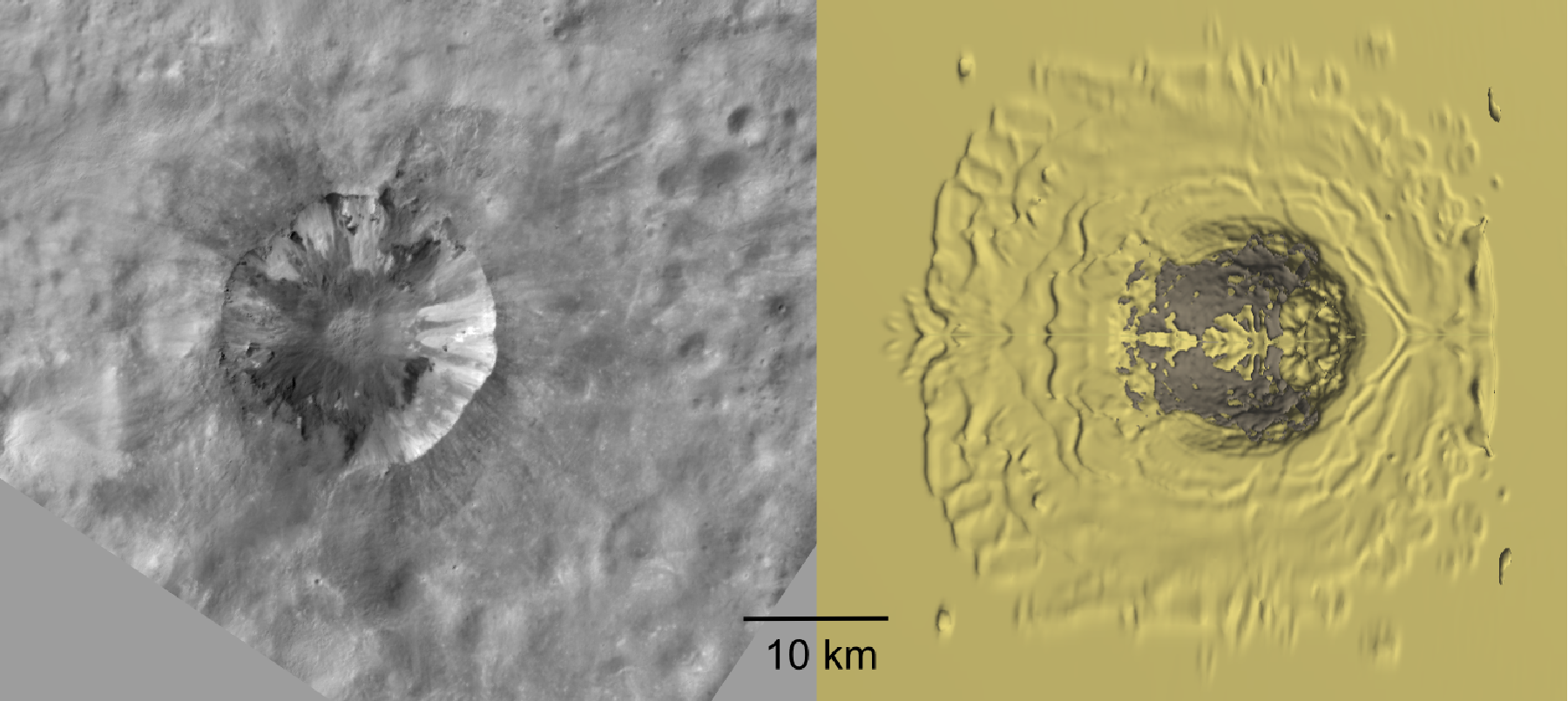}
\caption{{\color{black}Image of the Cornelia crater taken by the Framing Camera on-board the Dawn spacecraft during the HAMO phase (left panel) and the snapshot of the related hydrocode simulation, showing the distribution of the impactor's material inside the crater (right panel). The material from the impactor is shown in dark gray color, while the target material is shown in yellow color.
The impact direction is from the right and cuts the figure horizontally in the middle. The bilobate distribution of the projectile's material is symmetrical respect to the impact direction. The simulation also reproduced correctly the region devoid of dark material extending upstream from the crater center and symmetric respect to the impact direction. The main difference between the results of the simulations and the dark material inside Cornelia are due to a series of landslides that are identifiable as a set of mostly circular, brighter regions among the dark material near the top of the crater walls.}}\label{fig-cornelia}
\end{figure*}

{\color{black}
\subsection{Continuous flux, stochastic events and micrometeoritic flux}\label{results-scenarios}
%{Comparing the ``continuous flux'', the ``stochastic events'' and the ``micrometeoritic flux'' hypotheses}

As we mentioned in Sect. \ref{results-uncertainty}, the comparison of Figs. \ref{nflux} and \ref{mflux} shows that the delivery of the exogenous materials is dominated by the contribution of the largest impactors. The $1-3$ largest impact events of each class of impactors (i.e. dark and non-dark) deliver in fact about $30\%$ of the whole budget of the respective exogenous material over $3.98$ Ga. If we consider instead the largest $\sim6-8$ impact events, the delivered fraction of the relevant exogenous material increases to about $50\%$.

The comparison between Figs. \ref{nflux} and \ref{mflux} also shows us something else. Extrapolating from the results of \citet{bottke2005b} for the size-frequency distribution of the asteroid belt at sub-km diameters, the increasing trend in the number of impacts for decreasing diameters of the impactors should continue and Vesta should have received $\sim10^{6}$ impacts of decimetre-sized particles over the last $3.98$ Ga. According to the results of \citet{gladman2009}, however, the decreasing trend in the mass contribution to the exogenous material for decreasing diameters of the impactors should continue, as a reflection of the differential slope being shallower than $-3$ as discussed in Sect. \ref{methods-population}.

These considerations should immediately confirm that the ``micrometeoritic flux'' and the ``stochastic events'' hypotheses \citep{reddy2012,desanctis2012} are nothing more than two aspects of the ``continuous flux'' scenario. Micrometeorites did contribute to the contamination of Vesta over the last $\approx4$ Ga but, as also pointed out by \citet{reddy2012}, their contribution was not the dominant one. The largest impactors, on the other hand, did contribute a significant fraction of the dark material. However, to consider only their contribution would underestimate the budget of the exogenous material by a factor $2-3$ and would not explain the observed number of dark craters on Vesta, especially those inside Rheasilvia.
}

{\color{black}
\subsection{Cratering erosion of Vesta and the removal of the exogenous material}\label{results-erosion}

Now that our model is reasonably validated for what concerns the choice of the impact parameters and the predictions on the flux of impactors, we can use it to explore the contamination history of Vesta. We will start by assessing how old the dark material on the surface of the asteroid can be and how reasonable our choice of focusing on the post-Late Heavy Bombardment phase is. We will first investigate the role of cratering erosion, while in Sect. \ref{results-blanketing} we will explore that of crater saturation and ejecta blanketing.

The erosion of Vesta over the last $3.98$ Ga, estimated using Eq. \ref{erosion},  amounts to about $8.2\times10^{19}$ g ($0.03\%$ of the present vestan mass). This is equivalent to the loss of about $30$ m of material extending outward from the present surface of the asteroid. Over the last $1$ Ga, the mass loss amounts instead to about $1.3\times10^{19}$ g ($0.005\%$ of the present vestan mass) or, equivalently, to the erosion of about $5$ m of material.

Using Eq. \ref{erosion} with the sizes of the impactors that \citet{jutzi2013} assumed to have formed Veneneia and Rheasilvia, we can estimate that the mass loss associated with the two basins would be of the order of $10^{21}$ g ($\sim0.4\%$ of the present vestan mass). The erosion of the asteroid is therefore dominated, over the last $4$ Ga, by these two impact events. Note, however, that the match between the results obtained by \citet{jutzi2013} and the compositional signatures of the Rheasilvia basin is still matter of debate \citep{mcsween2013}, so the values we provided should be considered only for comparison purposes.

Even ignoring the roles of Veneneia and Rheasilvia, however, the balance between mass loss and mass gain is markedly in favour of the former. If, over a certain temporal interval, impacts were numerous enough to saturate the surface of Vesta (as we will discuss in Sect. \ref{results-blanketing}), the erosive effects associated with cratering could have removed a significant fraction of the previously deposited exogenous material. 
}

{\color{black}
\subsection{Crater saturation and blanketing effects}\label{results-blanketing}

As we mentioned previously, from the point of view of the observations of the Dawn spacecraft, the blanketing due to the ejecta produced by impacts of non-dark impactors counteracts the deposition of dark material by the dark impactors. Moreover, as we discussed in Sect. \ref{results-erosion}, all impacts contribute to the erosion of the surface of Vesta and, therefore, to the removal of any previously deposited exogenous material.

{\color{black}
In Figure \ref{fig-rplot} we show the crater populations produced by impacts on Vesta over the last $3.98$ Ga and the last $1$ Ga. In the plot we show also the  $5\%$ and $13\%$ saturation levels: these two threshold levels represent respectively the minimum value for which a crater population can reach equilibrium \citep{melosh1989} and the value estimated for Mimas, whose surface is the most densely cratered in the Solar System \citep{melosh1989}. As can be immediately seen, the crater population produced over the last $1$ Ga does not cause saturation effects. The impacts that Vesta received over the last $3.98$ Ga, however, produce saturation levels comprised between $5\%$ and $13\%$ for crater diameters between $50$ km and $100$ km, and reach the $13\%$ saturation level between $10$ km and $50$ km. 

As discussed by \citet{marchi2012}, the number of craters retained on the surface of Vesta will be lower than that of the craters produced (see Fig. S2 and the associated discussion in the Supplementary Materials of \citealt{marchi2012}). The values we obtained for the crater production over the last $3.98$ Ga are in broad agreement with the theoretical curves computed by \citet{marchi2012} for old terrains (see Fig. S2 in the Supplementary Materials of \citealt{marchi2012}), which in turn are in agreement with the crater population of the oldest terrains on Vesta, once corrected to obtain the retained population \citep{marchi2012}. Our results then imply that the mass loss due to cratering erosion discussed in Sect. \ref{results-erosion} globally affected the surface of the asteroid.
}
The effects of the Late Heavy Bombardment in terms of crater production are equivalent to another $1$ Ga of collisional evolution ($\sim300$ impacts, S. Pirani, Master Thesis at the University of Rome ``La Sapienza''; see also \citealt{turrini2013}). Once these craters and the two southern basins Rheasilvia and Veneneia are added to those produced in the following $3.98$ Ga, it would appear that the crater record on the surface of Vesta cannot allow us to probe much earlier than the Late Heavy Bombardment itself (as suggested also by \citealt{turrini2013}). Given that impacts remove more material than they deposit, it is therefore likely that the exogenous material delivered to Vesta during its first $0.5$ Ga of life was mostly stripped off by later impacts. In this case, the bulk of the dark material we see today on Vesta was brought on the asteroid after the Late Heavy Bombardment, justifying our initial choice of focusing on the post-Late Heavy Bombardment timespan.

Even when the crater production rate becomes low enough not to cause saturation effects, ejecta blanketing can still influence the amount of the dark material that reside on the surface of Vesta by burying it at different depths. We therefore computed the extension globally covered by ejecta over the two temporal intervals we focused on. The results are summarised in Table \ref{blanketing}, where we can see that the cumulative amount of ejecta produced over $3.98$ Ga would be enough to cover $1.7$ times the surface of Vesta. The fact that part of the dark material deposited in the last $3.98$ Ga should be buried by later ejecta provides a natural explanation to the dark veneers observed in crater walls by \citet{jaumann2012}.
The ejecta produced over the last $1$ Ga, on the contrary, would affect only $36\%$ of the vestan surface. 

We used our model to estimate what could be the plausible age of the dark material currently visible on the surface of Vesta. Note that here we are referring only to the dark material that is observed by the Framing Camera and the VIR spectrometer on-board the Dawn spacecraft, which are sensitive to the topmost $\sim1$~cm of the vestan regolith. The contaminants mixed in the topmost $1$~m of the vestan regolith are instead measured by GRaND and will be discussed in Sect. \ref{results-dark_contaminants} and Sect. \ref{results-bright_contaminants}. As we showed in Table \ref{blanketing}, impacts over the last $2.6$ Ga produce enough ejecta to cover a surface equal to that of Vesta. The dark material deposited from $2.6$ Ga ago to now should, on average, remain exposed on the surface of the asteroid, confirming the prediction we made in the Supplementary Information of \citet{mccord2012}. The surface coverage due to the dark material younger than $2.6$ Ga should be of the order of $14\%$ of the vestan surface. 

The values of the surface coverage here estimated should be corrected, however, for the still undetermined effects of the fraction of buried dark material that was exposed by later impacts and of the dark material mixed in the regolith by the micrometeoritic flux. Nevertheless, in terms of orders of magnitude the darkened surface visible to date on Vesta should vary between $\sim10\%$, which is the same order of magnitude of the darkest regions in Figure \ref{reflectance-map} ($R < 0.26$ or, equivalently, up to the magenta region), and $\sim30\%$, which is the same value estimated by GRaND for the extension of the most H-rich (H concentration equal to or higher than $250$ $\mu$g/g, \citealt{prettyman2012}) regions. 

{\color{black}
Note that, if we compute the crater population produced over the last $2.6$ Ga and plot the results in the same way as Figure \ref{fig-rplot}, the population of craters below $\sim$$70$ km 
%$$
will surpass the $5\%$ saturation level. It is therefore possible that craters produced over the last $2.6$ Ga removed part of the dark material already on the surface. This could explain the existence of the dark spots not associated with craters observed  on Vesta \citep{reddy2012}. The relevant craters could have been removed by later impacts leaving only the dark material deposited outside their borders, e.g. in the downstream region respect to the impact direction \citep{artemieva2011}.
}

\begin{figure}[h]
\centering
\includegraphics[width=\columnwidth]{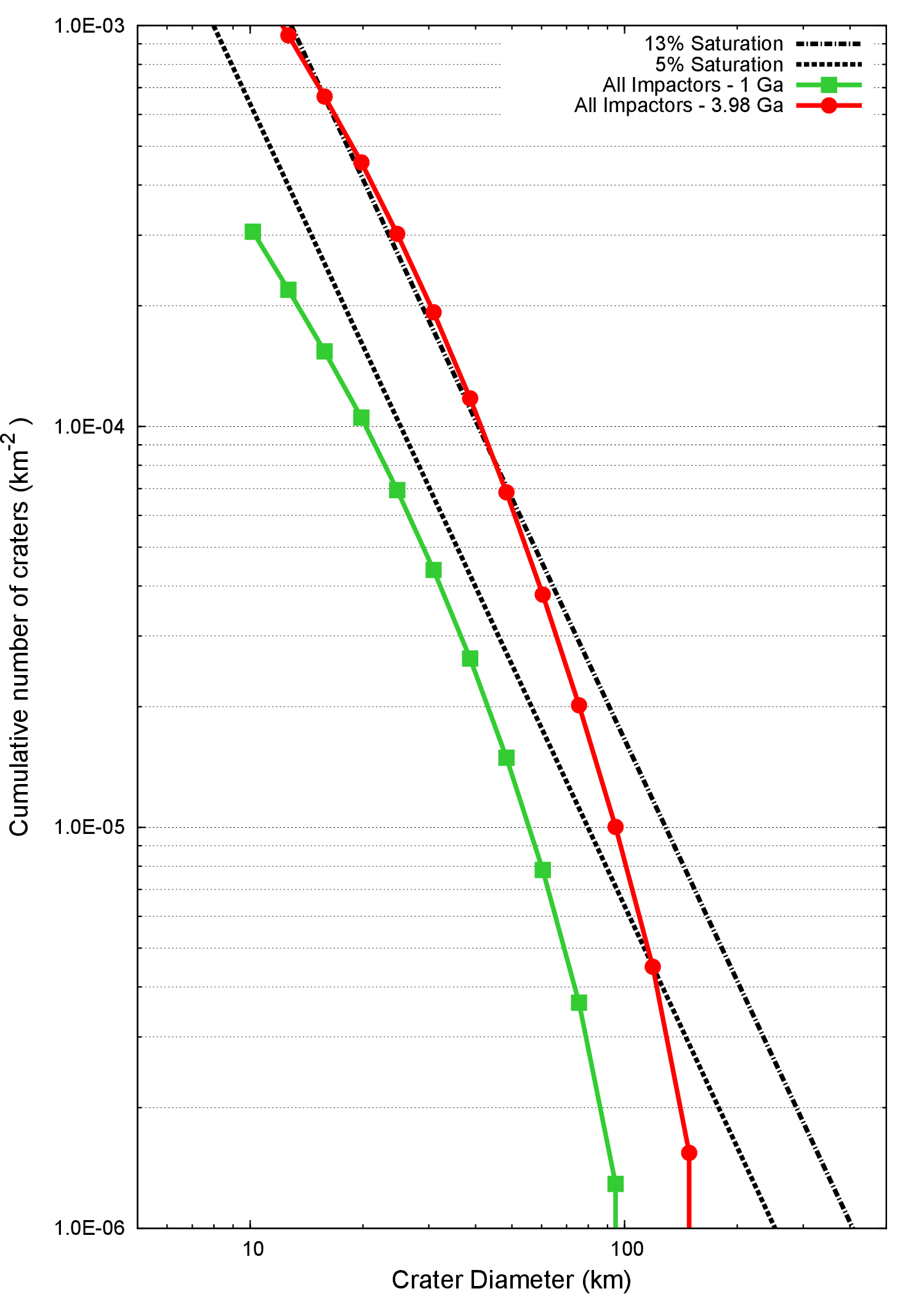}
\caption{{\color{black}Size-frequency distribution of the crater population produced on Vesta by all impactors over the last $3.98$ Ga and the last $1$ Ga. The $5\%$ and $13\%$ saturation levels are shown for reference. The effects of Veneneia and Rheasilvia are not included in this plot.}}\label{fig-rplot}%{fig-masses}
\end{figure}
}

\begin{table*}[t]
\begin{center}
\begin{tabular}{ccc}
\hline
%{} & \multicolumn{3}{c}{\bf{N. of Impacts}} \T\B\\
%\hline
%\bf{Impactors} &
\multirow{2}{*}{\parbox{0.3\linewidth}{\centering \head{Temporal Interval}}} & \multirow{2}{*}{\parbox{0.3\linewidth}{\centering \head{Non-Dark Blanketing (units of vestan surface)}}} & \multirow{2}{*}{\parbox{0.3\linewidth}{\centering \head{Dark Blanketing (units of vestan surface)}}} \T\B\\
\\
\hline 
$3.98$ Ga & $1.70$ & $0.25$ \T \\
$2.20$ Ga & $1.00$ & $0.14$ \T\B\\
$1.00$ Ga & $0.36$ & $0.04$ \B \\
\hline
\end{tabular}
\caption{{\color{black} Amounts of the vestan surface affected by the ejecta blanketing due to all impactors (Eq. \ref{blanketed}) and darkened by the fragments of the dark impactors (Eq. \ref{darkened}) over the two temporal intervals considered in this work. Also shown is the age ($2.6$ Ga) for which the blanketed area becomes equal to the surface area of Vesta. The quantities are expressed in units of the surface area of Vesta.}}\label{blanketing}
\end{center}
\end{table*}

{\color{black}
\subsection{Contamination by dark and water-equivalent materials}\label{results-dark_contaminants}

%\begin{table}
%\begin{center}
%\begin{tabular}{ccc}
%\hline
%\bf{Temporal} & \bf{Dark Material}& \bf{H$_{2}$O} \T \\%& \bf{Fe}  & \bf{Ni}  \T \\
%\bf{Interval} & \bf{(g)}          & \bf{(g)}      \B \\%& \bf{(g)} & \bf{(g)} \B \\
%\hline
%\multicolumn{3}{c}{\bf{Svetsov}} \T\\
%$3.98$ Ga & $3.43\times10^{18}$ & $8.17\times10^{16}$ \T \\
%$1.00$ Ga & $3.84\times10^{17}$ & $9.17\times10^{15}$ \B \\
%\multicolumn{3}{c}{\bf{Svetsov-Ong}} \T\\
%$3.98$ Ga & $1.16\times10^{18}$ & $2.42\times10^{16}$ \T \\
%$1.00$ Ga & $1.14\times10^{17}$ & $2.72\times10^{15}$ \B \\
%\hline
%\end{tabular}
%\caption{Masses of dark material and water-equivalent material delivered on Vesta in $3.98$ Ga and $1$ Ga by low albedo impactors using Eq. \ref{retention_svetsov} (label ``Svetsov'') and Eq. \ref{retention_ong} (label ``Svetsov-Ong'') to determine the retention efficiency of the asteroid.} \label{water}
%\end{center}
%\end{table}

\begin{figure}
\centering
\includegraphics[width=\columnwidth]{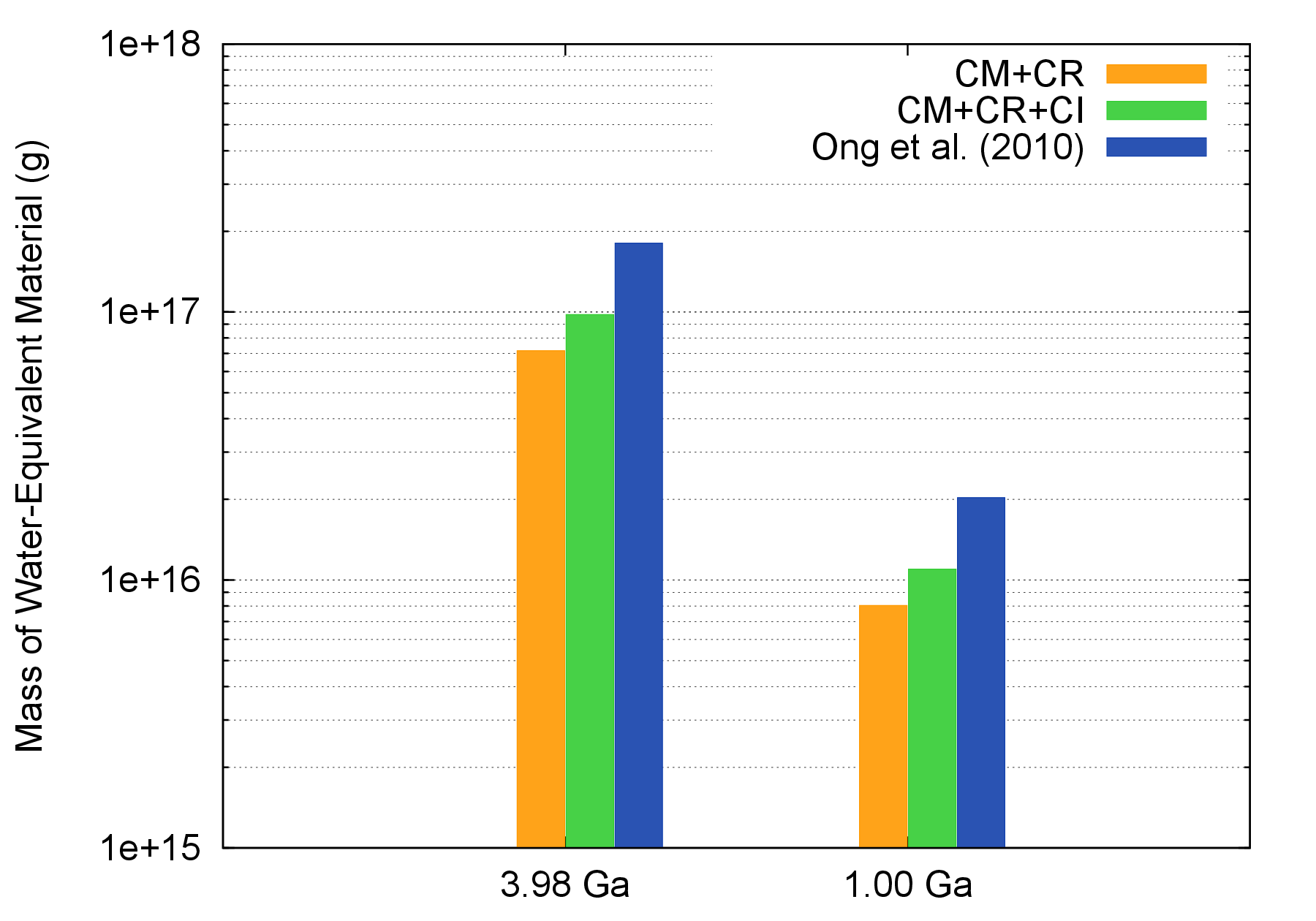}
\caption{{\color{black}Masses (in g.) of water-equivalent material delivered to Vesta over the last $3.98$ Ga and the last $1$ Ga by dark impactors for the different values of the fraction of potential water-carriers $F_{w}$ we described in Sect. \ref{method-dark_and_water}.}}\label{water-range}%{fig-masses}
\end{figure}

Since the previous results provided us a clearer picture of the survival of the dark material exposed on the vestan surface or buried in the vestan regolith, we can now proceed to quantitatively estimate the amounts of dark material and water-equivalent material delivered since the Late Heavy Bombardment and compare them with the measurements supplied by the GRaND instrument on-board the Dawn spacecraft.  

If we focus only on the size bins from Table \ref{sfd} that are associated with at least $1$ impact event respectively of dark, non-dark and all kind of impactors over the desired timespan (see e.g. Figure \ref{nflux} for the case of the last $3.98$ Ga), we can use the fluxes of impactors we computed to estimate the amount of the different exogenous materials delivered to Vesta (see e.g. Figure \ref{mflux} for the case of the last $3.98$ Ga). In this section we will focus only on the dark material and the associated water-equivalent material. We will discuss the cases of Fe and Ni in Sect. \ref{results-bright_contaminants}.
% We will extend the analysis of other kind of contaminants in the following section.

As shown in Table \ref{contamination}, over the last $3.98$ Ga Vesta would receive between $5.2\times10^{17}$ g and $1.8\times10^{18}$ g of dark material according, respectively, to Eqs. \ref{retention_ong} and \ref{retention_svetsov}. Over the last $1$ Ga, the amount would range between $5.8\times10^{16}$ g and $2.0\times10^{17}$ g again according, respectively, to Eqs. \ref{retention_ong} and \ref{retention_svetsov}. As the masses computed using Eqs. \ref{retention_svetsov} and \ref{retention_ong} always differ by a constant $3.4$ factor, in the following we will use the values given by Eq. \ref{retention_svetsov} as our guideline and will discuss the implications of a lower retention efficiency as the one given by Eq. \ref{retention_ong} only for the final results.

The delivered amounts of dark material previously discussed would be associated with the delivery of about $7.2\times10^{16}$ g of water-equivalent material over the last $3.98$ Ga and of about $8.0\times10^{15}$ g over the last 1 Ga, as shown in Table \ref{contamination}. As we show in Figure \ref{water-range}, the inclusion of CI chondrites among the carriers of water-equivalent material to Vesta would result in amounts  $1.4$ times as large. If instead we consider the water delivery rate assumed by \citet{ong2010}, the amount of water-equivalent material retained by Vesta would be about $2.5-3$ times larger than our reference case. In the following, we will focus on our reference case as it supplies the most conservative estimate of the amount of water-equivalent material.
}}

\begin{table*}
\begin{center}
\begin{tabular}{ccccccc}
\hline
\bf{Temporal} & \bf{Dark} & \bf{H$_{2}$O} & \bf{Non-Dark}     & \bf{Exogenous}    & \bf{Fe}  & \bf{Ni}  \T \\
\bf{Interval} & \bf{Material (g)}           & \bf{(g)}      & \bf{Material (g)} & \bf{Material (g)} & \bf{(g)} & \bf{(g)} \B \\
\hline
\multicolumn{7}{c}{\bf{Svetsov}} \T\B\\
$3.98$ Ga & $1.8\times10^{18}$ & $7.2\times10^{16}$ & $1.7\times10^{19}$ & $1.9\times10^{19}$ & $3.8\times10^{18}$ & $1.9\times10^{17}$ \T \\
$1.00$ Ga & $2.0\times10^{17}$ & $8.0\times10^{15}$ & $2.3\times10^{18}$ & $2.5\times10^{18}$ & $5.0\times10^{17}$ & $2.5\times10^{16}$ \B \\
\multicolumn{7}{c}{\bf{Svetsov-Ong}} \T\B\\
$3.98$ Ga & $5.2\times10^{17}$ & $2.1\times10^{16}$ & $5.1\times10^{18}$ & $5.6\times10^{18}$ & $1.1\times10^{18}$ & $5.6\times10^{16}$ \T \\
$1.00$ Ga & $5.8\times10^{16}$ & $2.4\times10^{15}$ & $6.8\times10^{17}$ & $7.4\times10^{17}$ & $1.5\times10^{17}$ & $7.4\times10^{15}$ \B \\
\hline
\end{tabular}
\caption{{\color{black}Masses of the dark (e.g. carbonaceous chondrites), non-dark (e.g. ordinary chondrites) and both exogenous materials, and masses of water-equivalent material (labelled ``H$_{2}$O''), Fe and Ni delivered to Vesta over the last $3.98$ Ga and $1$ Ga using Eq. \ref{retention_svetsov} (label ``Svetsov'') and Eq. \ref{retention_ong} (label ``Svetsov-Ong'') to determine the retention efficiency of the asteroid.}} \label{contamination}
\end{center}
\end{table*}

The GRaND detector is sensitive to the composition of the top-most $\sim1$ m of the surface of Vesta. The water-equivalent material estimated by GRaND should therefore represent a fraction of the total amount of the water-equivalent material mixed in the vestan regolith. The exact relationship between the value estimated by GRaND and the total amount of water-equivalent material depend on the unknown degree of its mixing in the vestan regolith and its uniformity with depth. From the H measurements of GRaND, \citet{prettyman2012} estimated a globally-averaged minimum content of $2.7\times10^{15}$ g of equivalent H$_2$O per meter depth of regolith. The nature of this lower bound, to first order, results in the removal of the post-Rheasilvia contribution of exogenous material. 

{\color{black}
The lower bound measured by \citet{prettyman2012} is in agreement with the range of values we reported in Table \ref{contamination} and in Figure \ref{water-range}. After removing the contribution of the post-Rheasilvia water-equivalent material, our model gives us an amount of water-equivalent material between $6.4\times10^{16}$ g. and $1.9\times10^{16}$ g. This would imply that the water-equivalent material present in the top-most $1$ m of the vestan regolith accounts for $4-15\%$ of the total budget of water-equivalent material, i.e. the regolith is not necessarily uniformly mixed but present a gradient of increasing concentrations moving toward the surface.  

This is consistent with the global picture obtained by our model once we take into account the results on the ejecta blanketing (see Table \ref{blanketing} and Sect. \ref{results-blanketing}). The dark material deposited before $2.6$ Ga ago was likely buried at different depths by the ejecta produced by subsequent impacts, thus explaining the veneers of dark material exposed at different depths in the walls of craters \citep{jaumann2012}. The dark material deposited after $2.6$ Ga, on the contrary, likely remained nearer to the surface and was mixed into a more limited amount of regolith by small-scale or micrometeoritic impacts. 

When we consider only the contribution of the last $2.6$ Ga, the amount of water-equivalent material delivered would range between $3.6\times10^{16}$ g. and $1.1\times10^{16}$ g. Once we subtract the post-Rheasilvia contribution, the presence in the topmost $1$ m of regolith of $10-30\%$ of the water-equivalent material delivered between $2.6$ Ga ago and $1.0$ Ga ago would be enough to explain the measurements of GRaND, leaving us enough margin to account for the removal of exogenous material due to the formation of Veneneia and Rheasilvia (not considered in this model). 
}

{\color{black}
\subsection{Contamination by Fe and Ni}\label{results-bright_contaminants}

Based on the agreement between our model and the observations of the Dawn spacecraft for what it concerns the flux of dark impactors and the delivery of water-equivalent material, we used our model to assess the delivery of Fe and Ni on Vesta by impacts and its implications for the composition of the vestan regolith. The results we obtained and that we will now discuss are summarized in Table \ref{contamination}. %As chondrites (both carbonaceous and ordinary) have, to first order, fairly similar Fe and Ni contents, the amounts of these elements delivered to Vesta scale linearly with the total amount of exogenous material.
Over the last $3.98$ Ga, Vesta should have received between $1.1\times10^{18}$ g and $3.8\times10^{18}$ g of Fe and between $5.6\times10^{16}$ g and $1.9\times10^{17}$ g of Ni. Over the last $1$ Ga, the exogenous Fe should vary between $1.5\times10^{17}$ g and $5.0\times10^{17}$ g and the exogenous Ni between $7.4\times10^{15}$ g and $2.5\times10^{16}$ g.

The Fe content of the vestan regolith has been constrained by the observations of GRaND \citep{prettyman2012,yamashita2013}, which showed that the average value of Fe/O and Fe/Si \citep{prettyman2012} and the range of values of Fe mapped by \citet{yamashita2013} are consistent with the values characteristic of HED meteorites (see \citealt{prettyman2012}, Supplementary Information). In particular, the composition of the vestan surface proved consistent, on the whole, with that of howardites \citep{prettyman2012,yamashita2013}.

To understand the implications of the Fe delivery for the composition of the vestan regolith, we performed the following tests. As our first test, we considered a layer of $100$ m of regolith (the minimum thickness according to present estimates) with an average density $\rho_{reg}=2000$ kg/m$^{3}$, whose total mass is $1.7\times10^{20}$ g. If we assume that the layer is uniformly mixed and we add to its natural Fe content (assumed to be $13.8$ wt$\%$ by  \citealt{yamashita2013} and in this work) the Fe contribution of all the exogenous material delivered over the last $3.98$ Ga (see Table \ref{contamination}), we get maximum Fe contents of $13.9-14.3$ wt$\%$.

As our second test, we followed our results on the contamination by water-equivalent material and focused only on the topmost $1$ m of the regolith, keeping the same density and average Fe content as in our previous test. The resulting mass of this $1$~m layer is therefore $1.7\times10^{18}$ g. We computed the possible amounts of Fe delivered to Vesta between $2.6$ Ga ago and $1$ Ga ago and we assumed that $10-30\%$ of this Fe (i.e. the same fractions as the water-equivalent material) was mixed in this regolith layer. The maximum Fe content we obtained this way is $15.7$ wt$\%$. The contribution of the following $1$ Ga, ignoring the impactors falling inside Rheasilvia as its contamination has been set back to zero by the formation of the basin, would raise the Fe content of the oldest terrains to about $16\%$.

In the first test, the values we obtain are well inside the range of the measurements performed by GRaND and are within $1\sigma$ ($\pm10\%$, \citealt{prettyman2012}) from their average value \citep{prettyman2012,yamashita2013}. In the second test, while our values fall outside the range of the measurements of GRaND \citep{yamashita2013}, they nevertheless fall inside $2\sigma$ from the average Fe content measured by GRaND \citep{prettyman2012}. Moreover, the variations caused by the exogenous Fe with respect to the ranges of possible native Fe contents of eucrites, diogenites and howardites were limited \citep{prettyman2012,yamashita2013}. Our results are therefore consistent with the observational data and indicate that Fe is not a reliable tracer of the contamination of Vesta.% as its effects could be mimicked by e.g. a higher concentration of basaltic eucrites in the regolith. 
%and are also in agreement with the fact that the Fe content of a pure 2:1 mixture of eucrites and diogenites and the Fe content of howardites are, as we discussed previously, virtually identical.

Differently from the case of Fe, the possibility to investigate the vestan Ni content with GRaND is still under evaluation. As a consequence, at present we can test the results of our model only against the HED meteorites. We did, nevertheless, perform the same tests as for the case of Fe. In the first test, we added the exogenous Ni from Tab \ref{contamination} to our $100$ m thick layer of regolith and obtained an average Ni content varying between $300$ $\mu$g/g (using Eq. \ref{retention_ong} for the retention efficiency) and $1100$ $\mu$g/g (using Eq. \ref{retention_svetsov} for the retention efficiency). This range is remarkably similar to that covered by the most Ni-rich howardites among the samples of \citet{warren2009}. It is also noteworthy that our range of values is practically identical to that of the samples that \citet{warren2009} label as ``regolithic howardites.'' In the second test, if we add the contribution of the exogenous Ni delivered between $2.6$ Ga and $1$ Ga ago to the native Ni content of the topmost $1$ m of regolith, we obtain that the Ni concentration can reach values up to $3600$ $\mu$g/g. Interestingly, the Ni content we obtain this way is of the same order of magnitude as the most Ni-enriched sample of howardites from \citet{warren2009}.
}

\section{Discussion and conclusions}
{\color{black}

In this work we explored the scenario of a continuous delivery of exogenous material to Vesta due to asteroidal impacts and we investigated its implications for the surface composition. We developed a model for the contamination of Vesta based on the one we originally used in \citet{mccord2012} to estimate the amount of dark material secularly accreted by the asteroid. We took advantage of the observations of the Dawn mission and of the laboratory data on the HED meteorites to verify our model in a quantitative way and, in turn, we used our model to derive possible observations that can help us disentangle the collisional and contamination history of Vesta.

The contamination scenario based on the continuous flux of exogenous material resulting from the secular collisional history of Vesta is one of the three explanations invoked for the dark material \citep{mccord2012} and the associated OH \citep{desanctis2012} and H \citep{prettyman2012} observed by the Dawn spacecraft. The two alternative explanations propose a dominant role either of the micrometeoritic flux, particularly at the time of the Late Heavy Bombardment, or of stochastic low-velocity impacts with large asteroids. 
Considering these three explanations as separate mechanisms is however misleading. As we showed with the help of our model, the ``micrometeoritic flux'' and the ``stochastic events'' scenarios are actually the end-members of the ``continuous flux'' scenario. 

The present day size-frequency distribution of the asteroid belt implies in fact that sub-km asteroids and micrometeorites represent the bulk of the impactors on Vesta, but their contribution in mass to the contamination is limited (as recognized also by \citealt{reddy2012}). Nevertheless, they have an important role in the mixing of the regolith and in determining the distribution of the contaminants with depth, an aspect that needs to be explored in future studies in order to improve the comparison of theoretical prediction with the data supplied by GRaND on the abundances of the different materials in the topmost $1$ m of the vestan regolith.
 
Large impactors, on the contrary, dominate the mass contribution to the contamination of Vesta and the $1-3$ largest impactors {\color{black}considered} in our model supply about $30\%$ of the total budget of contaminants. {\color{black}Stochastic events associated to even larger impactors, which we did not include in this analysis as the scaling laws we used do not necessarily hold for them and as their inclusion would require the calibration of our model with the complete crater record on Vesta, could bring amounts of exogenous contaminant comparable to the total budget of contaminants we estimated.
In particular, the impactors that caused the formation of Rheasilvia and Veneneia could easily have brought more exogenous material than all the other impactors cumulatively. They also likely removed or buried a significant fraction of the previously deposited contaminants. As to date models of their formation (see e.g. \citealt{jutzi2013}) do not yet reproduce satisfactorily the compositional signatures of the basins they created (see e.g. \citealt{mcsween2013} for a discussion), the role of Veneneia and Rheasilvia in the erosion and contamination histories of Vesta needs to be further investigated.}

\citet{reddy2012} and \citet{desanctis2012} also proposed that impactors before or across the Late Heavy Bombardment could have supplied a significant, if not the dominant, fraction of the dark material presently observed on Vesta by the Dawn mission. Vesta started collecting significant quantities of exogenous material immediately after the formation of its basaltic crust (see e.g. \citealt{turrini2011,turrini2013}) but, as soon as the impact velocities in the asteroid belt reached their present values (about $1$ Ma after the formation of Jupiter, {\color{black}\citealt{petit2002,turrini2011,turrini2012}}), Vesta started to lose more mass than the one it accreted \citep{turrini2011,turrini2013}. Therefore, the accumulation of the exogenous material delivered by the impactors was plausibly counteracted by the cratering erosion they caused.

As we showed with the help of our model, the crater population produced over the last $\sim4$ Ga can already produce saturation effects to a level between $5\%$ and $13\%$, in agreement with the observational data \citep{marchi2012}. This implies that cratering erosion affected the surface of Vesta to a global scale even after the Late Heavy Bombardment. Once we include the erosion associated with the formation of Rheasilvia and Veneneia to that due to the continuous flux of impactors, it seems likely that most material deposited before and during the Late Heavy Bombardment was removed from the surface of Vesta.

The results of our model highlighted that a full exploration of the link between the collisional and the contamination histories of Vesta over the last $4$ Ga is made difficult not only by the possibility of saturation effects among the crater population but also by the blanketing effects associated with crater production. The craters that formed on Vesta over the last $2.6$ Ga produced in principle enough ejecta to cover the whole surface of the asteroid. Over the same temporal interval, moreover, the impacts that formed Rheasilvia and Veneneia possibly caused a global blanketing by themselves \citep{jutzi2013}. These blanketing events make it difficult to assess the number of dark impactors that actually impacted Vesta before the formation of Rheasilvia. The effects of crater saturation and ejecta blanketing, on the other hand, explain in a natural way the dark spots not associated with specific craters \citep{reddy2012} and the veneers of dark material exposed on crater walls \citep{jaumann2012}.

While the ages of the oldest terrains on Vesta are still uncertain \citep{marchi2012}, the age of Rheasilvia basin is reasonably well constrained to about $1$ Ga \citep{marchi2012,schenk2012}. Such a young age implies that saturation and blanketing effects should not have affected the crater population that formed in its interior, therefore making it the best case study to calibrate our model. The fraction of dark asteroids in the asteroid belt is still quite uncertain and, to a lower extent, so is the number of dark features inside Rheasilvia. An updated count performed in the framework of this study revealed $16$ dark craters instead of the $12$ reported by \citet{reddy2012}. Still, in the limits of these uncertainties, all the abundances of dark asteroids that we considered in our model produced the correct order of magnitude of the dark craters observed in Rheasilvia, suggesting that dark asteroids represented about $20-30\%$ of the impactors on Vesta (at least during the last $1$ Ga).

The second calibration test that we performed focused on verifying whether the average impact velocity and impact angle we assumed for all impactors were appropriate to discuss the delivery of the dark material. We therefore simulated the formation of Cornelia crater and the distribution of the dark material associated with its interior \citep{reddy2012} using the iSALE-3D hydrocode. The results of our simulation reproduced reasonably well the bilobate distribution of the dark material inside the crater, the region devoid of dark material upstream to the crater center and their symmetry respect to the impact direction. The main differences between the results of the simulation and the dark material inside Cornelia are due to a series of landslides along the crater walls and near the top of the rim.%, easily identifiable as a set of mostly circular, brighter regions among the dark material near the top of the crater walls. 

As both previous tests produced a reasonable agreement with the observational data, we used our model to estimate the fraction of the vestan surface that should be affected by the dark material and the amounts of the different contaminants brought to Vesta since the Late Heavy Bombardment. Concerning the  darkening of the surface of Vesta, the comparison between the darkened area assessed from the Dawn's observations and the one estimated through our model produces a reasonable match in terms of orders of magnitude. Concerning instead the contaminants, we focused on three tracers: water-equivalent material (i.e. the amount of water needed to explain the observed OH and H on Vesta), Fe and Ni. For the first, the measurements of GRaND provided the distribution and a lower bound to the global abundance of the pre-Rheasilvia water-equivalent material. GRaND also provided the average global Fe content \citep{prettyman2012} and its regional distribution \citep{yamashita2013} in the vestan regolith, both consistent with the range of values characteristic of HED meteorites. The possibility of detecting Ni with GRaND, instead, is still under assessment, so our results can provide a guideline for future studies.

The amounts of water-equivalent material resulting from our model are consistent with the measurements of GRaND if about $5-10\%$ of the water-equivalent material delivered from the Late Heavy Bombardment to the formation of Rheasilvia remained mixed into the topmost $1$ m of the vestan regolith. When we focused only on the water-equivalent material delivered after the cumulative effects of ejecta blanketing became less important (i.e. from $2.6$ Ga ago to $1$ Ga ago), we noticed that the presence of about $10\%$ of the globally delivered water-equivalent material in the topmost $1$ m would be enough to match the measurements of GRaND. Once we consider the range of possible depths reached by the vestan regolith \citep{jaumann2012}, these results suggest that the dark material and the associated water-equivalent material should be present in decreasing quantities with depth, and a significant fraction should have remained in close proximity with the surface.
 
We took advantage of these results in our assessment of the implications of the delivery of exogenous Fe for the vestan regolith and we considered two scenarios. The first scenario assumed a uniform mixing of the exogenous Fe into a regolith layer whose thickness was set to the minimum value currently estimated through Dawn's observations \citep{jaumann2012}. The second scenario assumed instead that the distribution of Fe followed that of the water-equivalent material and resulted in a higher concentration in the topmost $1$ m of the vestan regolith. In both cases the effects of exogenous Fe for the total Fe abundance fell inside the range of the GRaND's uncertainty on the measurements and were globally negligible with respect to the intrinsic variability of the Fe content of HED meteorites. As a consequence, Fe did not prove to be a reliable tracer of the contamination by exogenous material.

Ni, on the contrary, proved to be an efficient tracer of the contamination: the amounts of Ni estimated to have been delivered by impactors over the last $\sim4$ Ga, if mixed uniformly into $100$ m of vestan regolith, would produce the range of values observed in those howardites that \citet{warren2009} label as ``regolithic howardites''. If, once again, we take advantage of the results obtained for the water-equivalent material and assume that $10\%$ of the exogenous Ni delivered since $2.6$ Ga ago is confined in the topmost $1$ m of the vestan regolith, we obtain the Ni content of the most Ni-enriched sample studied by \citet{warren2009}. As a consequence, should GRaND prove capable of detecting Ni in the vestan surface, the study of the Ni abundance and distribution could potentially be a powerful counterpart to dark and water-equivalent materials in unveiling the contamination history of Vesta.

As we pointed out, the model we developed is affected by two intrinsic sources of uncertainties. The first one is our limited knowledge of the real values of the parameters (e.g. the fraction of dark impactors in the asteroid belt) the model uses to characterize the collisional and contamination histories of Vesta. The second one is due to the {\color{black}statistical} nature of the evaluation of the collisional history. The largest impact events associated with the estimated fluxes of dark and non-dark impactors, due to their stochastic nature, may or may not have taken place. As these events can contribute {\color{black}} $20-30\%$ of the budget of the relevant exogenous material {\color{black}(and possibly increase it by a factor of a few)}, they significantly affect our estimates of the vestan contamination.

This second source of uncertainty can be circumvented in future investigations by calibrating the model over different temporal intervals (hence fluxes of impactors). This can be done by applying the model to selected areas of the vestan surface characterized by different ages and for which the crater records are well preserved. It must be noted that the improvement of the crater catalogue of Vesta, with its extension to the northern hemisphere and the identification of possible ancient degraded craters, will automatically provide more stringent constraints to the fluxes and the sizes of the impactors. Nevertheless, the model is already capable of reproducing the correct orders of magnitude of the observational and laboratory data available on Vesta and the HED meteorites. {\color{black}We also must emphasize that the implications of neglecting the effects of large, stochastic impacts (a choice that decreases the budget of contaminants in our analysis)  should be at least partially mitigated by our choice of considering the lowest limit of the thickness of the vestan regolith in which to mix the exogenous material (therefore magnifying the effects of the contaminants).}

The first source of uncertainty, however, is more difficult to overcome and has important implications, in particular for what concerns the fraction of dark asteroids in the asteroid belt. The calibration we performed on Rheasilvia seems to suggest that the fraction of dark impactors on Vesta over the last $1$ Ga ranged between $15-30\%$, in agreement with the estimated abundance of dark impactors done by \citet{desanctis2012}. The amount of water-equivalent material estimated using this fraction of dark impactors agrees with the lower bound estimated by GRaND.  However, in such a scenario about $70-85\%$ of the impactors on Vesta would be non-dark. The present collection of HED meteorites points instead in the opposite direction. Howardites and polymict eucrites have been shown to contain clasts of chondrules, metals, ordinary chondrites, enstatite chondrites and possibly achondrites \citep{lorenz2007} but the largest fraction (about $99\%$, \citealt{lorenz2007}) of the xenoliths in HED meteorites is due to carbonaceous chondrites \citep{zolensky1996,lorenz2007}. Presently, we cannot provide a conclusive answer to this conundrum, but we can however formulate some hypotheses on its possible causes.

The first, obvious possibility is that the fraction of dark impactors we are using is not the correct one. Interestingly enough, this has no implications for the amount of delivered Fe and Ni as, to first order, carbonaceous chondrites and ordinary chondrites contain the same fractions of these elements \citep{jarosewich1990}. The amount of water-equivalent material would instead increase together with the number of dark impactors. As GRaND provided a lower limit to the H abundance in the topmost $1$ m of the vestan regolith, increasing the delivered amount of water-equivalent material does not change the agreement between the model and the observational data. The only difference {\color{black}with} respect to our original discussion is that the water-equivalent material does not need to be more concentrated toward the surface but can be more uniformly mixed into the vestan regolith. 

The major implication of a larger fraction of dark impactors would be for the number of dark craters, especially those inside Rheasilvia. Increasing the fraction of dark impactors to $90\%$ of the total flux (i.e. about the same fraction of the carbonaceous clasts among the xenoliths in HED meteorites) would produce about $54\pm7$ dark craters inside the basin. The most updated count of the dark craters inside the basin is $16$, which is more than $3\sigma$ away from the estimated value. On the other hand, such a high frequency of dark impactors would deliver enough dark material and water-equivalent material to reproduce the measurements of the Dawn mission also in the case Veneneia and Rheasilvia globally covered the surface of Vesta with their ejecta blankets, burying most of the oldest contaminants to depths beyond the capabilities of the instruments on-board the Dawn spacecraft.

The second possibility is that the story the HED meteorites tell us about their contamination is not complete or is misleading. It is possible that the collection of samples that fell on the Earth is not representative of the whole surface of Vesta, being linked to one or a few specific cratering events (e.g. Rheasilvia, Veneneia or both). It is also possible that most HED meteorites are actually fragments of V-type asteroids, whose lower gravity respect to the vestan one would make it easier to excavate them. V-type asteroids mainly (but not exclusively, see e.g. \citealt{mcsween2011}, \citealt{obrien2011} and references therein) originated from Vesta, but their small sizes imply that their collisional histories are dominated by sub-km and plausibly sub-m sized impactors, contrary to the case of Vesta discussed in this work. Given that carbonaceous chondrites are composed by more fragile materials than ordinary chondrites (see e.g. \citealt{lorenz2007}), it is not implausible that their parent bodies are more easily fragmented or eroded \citep{lorenz2007} and that, as a consequence, carbonaceous chondrites represent the bulk of sub-km and micrometeoritic impactors on Vesta and the V-type asteroids. The dominant presence of carbonaceous chondrites in HED meteorites could therefore be due to a selection effect caused by the smaller sizes of the V-type asteroids. 

A third possibility is that, while our samples are indeed representative of Vesta alone, it is our understanding of the data that is not complete. The gravity of Vesta is strong enough that the average retention efficiency does not change dramatically across the range of plausible impact velocities, going from $42\%$ to $68\%$ between $2$ km/s and $10$ km/s. Therefore, a large fraction of material from the impactors should remain on the surface of Vesta almost independently of the impact velocity. Low impact velocities are generally invoked for the delivery of most of the dark material to explain its preserved hydrated state \citep{reddy2012}. The less abundant dehydrated clasts in HED meteorites are generally regarded as the product of impacts occurring at higher velocities \citep{reddy2012}. As pointed out by \citet{lorenz2007}, however, in HED breccias much of the chondritic material is contained in disseminated form. These authors argue that concentration of exogenous material revealed by the abundance of siderophile elements is significantly larger (sometimes by orders of magnitude) than that associated with the presence of clasts and impact melts, with the bulk of the siderophile elements present as sub-microscopic inclusions. Dark impactors could have therefore hit Vesta over the whole range of impact velocities, supplying contaminants in different form.

Ordinary chondrites are generally less fragile than carbonaceous ones, as showed by the fact that they survive more easily to the passage through the atmosphere of the Earth (see e.g. \citealt{lorenz2007}). It is therefore likely that they are better preserved during the impacts on Vesta, and it is conceivable that most of the mass of low-velocity, ordinary chondrite-like (i.e. non-dark) impactors could end up in boulders and fragments larger than those produced by carbonaceous chondrite-like impactors. These larger fragments could be less easily incorporated into the regolith, differently from the smaller fragments produced by faster or more fragile impactors. Most non-dark contaminants could therefore be present among the vestan regolith as discreet units (large rocks or boulder) instead of being diluted in the form of small fragments and clasts. This scenario would allow {\color{black}us} to fit the results of our model into all available observational and laboratory constraints, but dedicated, high-resolution impact simulations are needed to test whether this is a viable possibility or not.

{\color{black}Finally, it is also possible that, when we look at the data from the Dawn mission and at those from HED meteorites, we are simply looking at two different aspects of the same problem. The study of \citet{nesvorny2010} showed that the population of sub-millimetre particles in the near-ecliptic zodiacal dust is contributed only up to about $\sim$$10\%$ %$$
by asteroids while about $\sim$$90\%$ %$$
is due to Jupiter-family comets. \citet{nesvorny2010} proposed that these cometary dust particles are responsible for the micrometeoritic flux on the Earth, which is dominantly carbonaceous in composition (in contrast to the meteoritic flux, which as we previously said is dominated by ordinary chondrites). This implies that also the micrometeoritic flux on Vesta should be mainly composed by carbonaceous particles. As we discussed in this work, in the ``continuous flux'' scenario the micrometeorites dominates \textit{in number} the flux of impactors on Vesta, thus naturally explaining the 
carbonaceous composition of the almost totality of the xenoliths in HED meteorites \citep{zolensky1996,lorenz2007}. Conversely, the delivery of exogenous material on Vesta is dominated \textit{in mass} by  asteroids with diameters of several kilometres or bigger (with the largest impactors contributing most of the contaminants), whose contribution can correctly reproduce the measurements of the Dawn mission. Intermediate, km-sized (and sub-km) asteroids would contribute a fraction of the total budget of contaminants but would dominate the crater population of Vesta. In particular, dark impactors in this size range would naturally explain the dark craters observed inside the Rheasilvia basin by the Dawn mission. 
}

\section*{Acknowledgements}

{\color{black}The authors wish to thank the NASA Dawn project and the entire Dawn team for the development, cruise, orbital insertion, and operations of the Dawn spacecraft at Vesta, and in particular the VIR, GRaND and FC instrument teams on whose data this work is based. The authors gratefully acknowledge the developers of iSALE-3D, including Dirk Elbeshausen, Kai W{\"u}nnemann and Gareth Collins, and wish to thank David W. Mittlefehldt for his comments and the discussions that helped to improve the model, Sharon Uy for her assistance in the preparation of the manuscript, and the two reviewers for their comments that contributed to improve both the model and the manuscript.} This research has been supported by the Italian Space Agency (ASI) and by the International Space Science Institute (ISSI) in Bern through the International Teams 2012 project ``Vesta, the key to the origins of the Solar System'' (\url{www.issibern.ch/teams/originsolsys}). The computational resources used in this research have been supplied by INAF-IAPS through the project ``HPP - High Performance Planetology.''

\bibliographystyle{elsarticle-harv}
%%\bibliography{<your-bib-database>}



\end{document}